\documentclass[a4paper,onecolumn]{article}
\usepackage{amssymb}
\usepackage{graphicx}
\usepackage{amsmath}
\usepackage[normalem]{ulem}
\usepackage{color}
\usepackage{authblk}
\usepackage[dvips]{geometry}
\title{Detailed study of a moving average trading rule}

\author[1]{Fernando F. Ferreira\thanks{ferfff@usp.br}}
\author[2]{A. Christian Silva\thanks{csilva@idatafactory.com}}
\author[3]{Ju-Yi Yen\thanks{ju-yi.yen@uc.edu}}

\affil[1]{Center for Interdisciplinary Research on Complex Systems,
  Universidade de S\~{a}o Paulo,03828-000 S\~{a}o Paulo-SP, Brazil
  }
\affil[2]{IdataFactory, Houston, Texas 77030, USA}
\affil[3]{Department of Mathematical Sciences,
University of Cincinnati, Cincinnati, Ohio 45221-0025, USA
}

\begin{document}

\maketitle

\begin{abstract}
We present a detailed study of the performance of a trading rule that uses moving average of past returns to predict future returns on stock indexes. Our main goal is to link performance and the stochastic process of the traded asset. Our study reports short, medium and long term effects  by looking at the Sharpe ratio (SR). We calculate the Sharpe ratio of our trading rule as a function of the probability distribution function of the underlying traded asset and compare it with data. We show that if the performance is mainly due to presence of autocorrelation in the returns of the traded assets, the SR as a function of the portfolio formation period (look-back) is very different from performance due to the drift (average return). The SR shows that for look-back periods of a few months the investor is more likely to tap into autocorrelation. However, for look-back larger than few months, the drift of the asset becomes progressively more important. Finally, our empirical work reports a new long-term effect, namely oscillation of the SR and propose a non-stationary model to account for such oscillations.
\end{abstract}

\section{Introduction}

In this article, we use historical price data trading rules to predict future asset performance. We focus on the popular approach of using moving averages of returns. Important and well known examples of trading  rules that  use  historical  prices are short-term or long-term reversal rules \cite{Bondt1989}, cross-sectional momentum \cite{Antonacci,JT1993,Harvey}, time-series momentum \cite{Moskowitz2011} and trend-following \cite{BP,Papailias2015}. These strategies are characterized by the use of historical data to predict the  future.  In particular, short-term reversal is reported for portfolios built using monthly historical data. Trend-following and momentum are generally implemented using a few months to one year and long-term reversal uses a few years of historical data. Furthermore, this classification is based on equities where the vast majority of the literature  is  concentrated \cite{Harvey}. This study belongs to the extensive  literature of historical price based trading and presents an unifying view of all the different time frames (from short-term to long-term), generally  studied independently. 

Our work, however, differs from the majority of the literature in several aspects. First, we look at the performance of a single asset subject to a trading rule that uses past data of the asset itself, that is,  the technical rule of past performance, as in technical analysis \cite{Brock,Griffioen}. Second, our empirical work uses weekly stock indices and not stock prices. Third, we start our analysis by calculating first the closed solution form of the Sharpe ratio (SR) of our trading rule. We then link our solution to the stochastic process of the asset with the Sharpe ratio of our trading rule. Fourth, we are not concerned if such trading rules are true alpha opportunities. This point has been extensively debated in the literature \cite{Moskowitz2011,BP,Papailias2015,Harvey} where extensive statistics have been performed for trading rules similar to ours \cite{Moskowitz2011}. Our main goal is to link the performance of these rules with the underlying statistics of the traded asset. In particular, we show the effect of non-stationary data. Therefore, the nature of this study is explanatory and not predictive (just as \cite{shmueli2010}). Finally, our empirical work is presented graphically by taking the SR as a function of the moving average rolling window $N$ and showing that such an analysis is able to uncover the statistical nature of the performance.

When compared to the literature, this work is mostly an extension of the work of \cite{Moskowitz2011}  which  introduced  the  notion  of  time-series  momentum. Time-series momentum trades an asset based on the past trailing performance of the asset alone, in contrast to cross-sectional momentum \cite{JT1993}, where assets are traded relative to each other\footnote{
Cross-sectional momentum is not the focus of this article, but it has been studied extensively if compared to time-series momentum or trend-following. Since the work of  \cite{JT1993}, momentum has been extended to different asset classes, portfolios, and other markets abroad. Momentum has been reported in international equity markets by \cite{Doukas,Forner,Nijman,Rouwenhorst}; in industries by\cite{Lewellen,Moskowitz}; in indexes by  \cite{Bhojraj,Asness1997}; and in commodities by \cite{Erb,Moskowitz2011}. Single risky asset momentum is analyzed in \cite{Daniel,Barberis,Hong,Berk,Johnson,Ahn,Liu,Sagi} to cite a few. Recently, momentum has also been studied linking its performance to business cycles and regimes by \cite{Chordia,Kim2013,Griffin,Guidolin,Barroso}.}. Time-series momentum rules are very close in nature to trend-following rules. While trend-following rules may involve prices  \cite{BP,Papailias2015}, time-series momentum ones look at past price returns. In summary, time-series momentum can be equated with the study of the simplest trend following strategies, and therefore a building block for more complex strategies. Moskowitz, Ooi and Pedersen perform an extensive study using over 25 years of indices, futures and forward contracts data and show that every asset they have analyzed (58 in total) present statistically significant positive returns using time-series momentum. 

In contrast with Moskowitz, Ooi and Pedersen, we do not focus exclusively on the typical momentum portfolio formation period. We look at portfolio formation periods from one week to ten years. That allows us to look at short-term and long-term data effects.  Another distinction is that our empirical analysis focus on stock indices exclusively, while the  analysis they perform is over many asset classes using mostly traded future contracts data. We do not use future contracts but the underlying cash indices.

In contrast with earlier work we start by defining the mathematical trading rule which can   be solved exactly for both risk and return. We use the same rule first introduced by \cite{Lo1990} for the case of cross-sectional portfolios of individual stocks. We also used the same algorithm used by \cite{Lewellen} to study portfolio of industries and also evoked  by \cite{Moskowitz2011} to contrast time-series momentum with cross-sectional momentum. However, we do not restrict ourselves to the formula of the average performance of the trading rule as a function of the probability distribution of the underlying traded security. We also solve for the  standard  deviation of the trading rule assuming that the probability of the underlying returns is a Gaussian distribution with complex autocorrelation relation between the returns at different time lags. Therefore, we look at the risk adjusted performance measured by the Sharpe ratio (SR) as a function of the look-back used to construct the portfolio. The theoretical mean and the variance of a moving average trading rule was  also derived by \cite{Grebenkov} for an exponential moving average of assets modeled by an Ornstein-Uhlenbeck process with zero drift. The work of Grebenkov and Serror is less empirical than ours, even though the mathematical formulation is more general, since the authors go beyond the first two moments. Furthermore, their formulation is by construction mostly focused on the importance of the autocorrelation, ignoring the drift. On the other hand, we are very interested on the effect of the drift and show that the drift can be important to the final performance

We emphasize  that  our  work  links  the  moments  of  probability  distribution  of  the  traded  assets with the performance using our SR formula. We do not link our findings to behavioral finance  \cite{Daniel,Barberis,Hong,Berk,Johnson,Ahn,Liu,Sagi}.  In general,  theoretical  studies  based  on  behavioral  finance  suggest  the trading decision follows past performances and if  prices  overreact  or  underreact  to  available information.  For  instance,  \cite{Daniel}  propose  that  under-reaction  and  overreaction  are  consequences  of  investors'  overconfidence on  inside  information and  self-attribution  bias.  \cite{Barberis}  connect  overreaction of prices to investors' attitude towards a series of good or  bad  news,  and  under-reaction to their attitude  towards  information such  as earning announcements.  In  \cite{Hong} model, the investors are categorized into two groups, the news watchers and the momentum traders, which leads to under-reaction at short horizons and overreaction at long horizons.\footnote{Alternatively to under and overreaction \cite{Daniel,Barberis,Hong}, there are other causes that have been cited as possible explanation. \cite{Lewellen} suggests that the lead-lag cross-serial correlation should explain cross-sectional momentum.
 \cite{Conrad} point to the cross-sectional variation of asset returns. Chordia and Shivakumar and others \cite{Chordia,Kim2013,Griffin,Guidolin,Barroso} study business cycles and suggest that time-varying expected returns can explain  momentum.} Generally speaking, this set of studies describes investors as  Bayesian optimizers: the investor observes or receives information at each investment time period,  and  updates  his/her  investment  decision according to his/her belief. However  we  do not attempt to address such theories; we  refer the reader  \cite{Moskowitz2011} for a brief discussion.

We split our empirical work in two parts. First, we work in-sample with data that is transformed to be approximately stationary.  Such transformations  require  that  we  know  the  entire history and therefore are by definition in-sample and cannot be applied for trading. In summary, we normalize the returns with a local measure of the volatility to reduce volatility clustering and   to approximate the probability distribution to a Gaussian one. Next, we break the data into regimes that   are on average two years long. Such regimes are chosen statistically and therefore do not necessary conform to economic cycles but are selected from the data in order to present a well-defined and constant drift. Finally, we look at the autocorrelation of the returns to define two autocorrelation regimes. The first covers data prior to 1975, where the returns are mostly positively autocorrelated at short lags. The second covers data after 1975, where the autocorrelation is mostly negative.

Next we work out-of-sample. By out-of-sample we  mean that our trading decision at time $t$ uses data up to time $t-1$. There is no hindsight. That is, we employ rolling forecasting origin also   know
as time series cross-validation \cite{Hyndman}. In this article, out-of-sample does not mean that we are dealing with a traditional hold out period where we would test the performance of some parameters selection. In summary, the out-of-sample study conforms with  most of the literature on the subject, which apply a trading rule over a long history  \cite{Moskowitz2011,BP}. This strategy does not require future information. However, our empirical study is by definition non-stationary. We are  particularly interested on the effects of using non-stationary data and contrast our in-sample results with our out-of-sample results. Once again, we emphasize that our goal here is not to suggest or advocate a trading rule, but to understand its underlying mathematical law by linking it to the stochastic process of the traded asset. Therefore the tradeability (such as transaction costs and portfolio allocation) of a full strategy is not our focus.

Our stationary/in-sample study shows that both autocorrelation and mean drift of the random process of the traded asset are important in the final performance of the trading rule. In particular, for look-back periods up to four months, the most important effect is the autocorrelation. For look-back periods between four months to one year, the drift becomes progressively more important. Therefore, we identify two clear phases in the SR as a function of the portfolio formation period. One is marked by autocorrelation, the other by the drift. The drift phase is characterized      by an increase of SR as a function of the portfolio formation period while the autocorrelation phase signature is different and normally characterized by a decrease of the SR with the increase of the moving average range.

Our non-stationary/out-of-sample results agree with our in-sample study only approximately. The au- tocorrelation phase is still clearly present for short look-back windows. The drift effect is much     less pronounced and clearly shows a reversal after portfolio formation periods of one year. This reversal conforms with the literature on stocks and also the work of Moskowitz, Ooi and Pedersen. However, after the reversal we find that the performance picks up again around look-back periods of four years. In fact we report damped oscillations of the SR where regimes close to “momentum” and ”reversal” ones alternate. We believe that this empirical result is an original contribution of this article. It should be noted that “clear” periodic oscillations are rare in finance. We are only aware of Heston and Sadka’s seasonality results \cite{HS2008}.

The contrast of our in-sample and out-of-sample results leads us to believe that the main    cause of the difference is the change in the drift of the underlying market, that is, the very nature    of the stationary assumption for the drift. Finally, we propose a simple model for the market that assumes that the drift is not constant but follows a periodic function, that is, the average return for the    asset changes from positive to negative every few years (approximately every two years for the US market). This simple model is more in line with \cite{Chordia} than traditional behavioral theories. Chordia and Shivakumar argue that the time-varying expected returns due to business cycles are responsible for momentum. However, we emphasize that our analysis cannot discern between different theoretical models. We only state that time-varying expected returns can be important to explain our findings, especially considering that the alternative (long range autocorrelation) appears to be less likely.

In section \ref{sec:t_model}, we present our trading rule and find the theoretical average performance, the theoretical standard deviation and the Sharpe ratio. Furthermore, in section \ref{sec:t_model} we also perform simulation studies to validade our calculations. In section 3 we present empirical results. We first look at stationary data comparing it to our theoretical formulas and then at non- stationary data. Some formula derivations are given in appendix \ref{sec:app1}. Appendix \ref{sec:app2}, \ref{sec:app3} and \ref{sec:app4} complement the empirical analysis. In particular, Appendix \ref{sec:app3} extends the analysis to international indices and Appendix \ref{sec:app4} presents the analysis for different trading frequency.

\section{Model} 
\label{sec:t_model}

\subsection{Trading rule}

The trading rule intends to extract predictability of future price returns from past price returns. We define price return by the following expression:
\begin{equation} \label{eq:logR}
X_{t} = \ln (S_{t}/S_{t-1}), 
\end{equation} 

\noindent where $S_{t}$ is the price in period $t$.

In order  to  understand  return  based  strategies,  we  use  a  proxy  algorithm  that  should  represent the general characteristics of any given trend-following/time-series momentum rule. The term trend-following here means that we trade in the direction of the moving average of past returns. It does not mean that our moving average is restricted to looking back one year. The trading rule we use is:

\begin{equation} 
\left\{ \begin{array}{lll} m_{t-1}(N) =
\displaystyle\sum_{i=1}^{N} X(t-i)/N, & ~ & ~\\ \\ \mbox{if} ~ ~ m_{t-1}(N) > 0 ~~
\mbox{Buy, Long} \\ \\ \mbox{if}~~ m_{t-1}(N) < 0 ~~\mbox{
Sell,  Short}
\end{array}\right. 
\label{eq:strategy} 
\end{equation} 

\noindent where m is a simple moving average, and N is the look-back period used to calculate the moving average. This same set of rules  was  used  in  \cite{Asness2013,Kim,Moskowitz2011,Lewellen,Lo1990}, and other prior    studies\footnote{There is no guarantee that such an algorithm represents well trend-following strategies, however  we  defer the question of how to represent a class of strategies for future study. For now, we consider that our algorithm is able to capture the main mathematical features present in a general moving average based strategy.}.

From a practical point of view, $m_{t-1}$ can be thought as fraction of the wealth to be traded at the time
 $t-1$. Notice that since $m_{t-1}$ has a magnitude that is typically very close to zero, it means that in order to
trade we need to rescale $m_{t-1}$. That can be achieved, for instance, by rescaling $m_{t-1}$ by a measure
of the standard deviation (see section 3 below). Furthermore, our trading rule (\ref{eq:strategy}) presents dynamic leverage by construction. That is, we are implicitly moving money between the asset and a bank account. We assume that such bank account pays or receives zero interest rate, which means that we may borrow to invest. In practice, we could restrict high leverage but we ignore implementation issues in this article.

Notice that since the bank account does not pay interest we do not need to account for it when calculating the SR of our trading     rule.

\subsection{Risk and Return for stationary random variables}

\subsubsection{Stationary random variables}
\label{sec:stationaryProcess}

The strict definition of a stationary process is that the joint probability distribution of all random variables is invariant under time shifts or translations. Equivalently, the probability density depends only on the time difference since the time origin is not relevant \cite{Feller,Gardiner}. However, we use in this article the weak-sense stationary definition. In other words, one requires that the first moment and covariance     do not change in time. In particular, the variance exists and the covariance depends only on the    time difference.

We take the view that financial time series are not weak-sense stationary in general. This includes many of the common transformation of the price time series, such as log-returns. Most of the literature does not discuss the effect of non-stationary data; however, some studies show measurable and important effects \cite{Seemann,McCauley}.

We follow \cite{Seemann,Silva,Tsiakas} and assume that the data has patches or regimes of stationary periods. In general, we do not postulate a periodic structure, nevertheless, we expect the presence of some repetition in order to have enough data to perform ensemble averages. Furthermore, in order to detect practically such periods, we shall simplify the problem by detecting periods with constant mean and constant variance. We assume that the autocorrelation is well-behaved enough to preserve the time independent property.

To detect periods with constant mean (drift), we use the “Breaks for Additive Season and Trend” (BFAST) algorithm \cite{Verbesselt}. BFAST begins by using the Loess regression to decompose the time series into seasonal trends and irregular components. Thereafter, the algorithm performs a loop where it uses the methodology developed by \cite{Bai} until the number and the position of the breakpoints are unchanged. Intuitively, the algorithm finds the trend by fitting a piecewise linearity where the breakpoints (changes from one linearity to the next) are   found at the same time to the linear fit. From an implementation standpoint, we used the R package “bfast” in our empirical studies.

Besides BFAST,  there  are  several  other  algorithms \cite{Matteson,Fukuda,McCauley,Seemann,Guidolin} but we postpone the comparison for a  future  study.

\subsubsection{Risk and Return}

The average return for our trading rule (Eq. (\ref{eq:strategy})) is given by

\begin{equation}\label{eq:R} 
\left \langle R \right \rangle = \frac{1}{T-N+1}
\sum_{t=N}^{T} m_{t-1}(N) X_{t} 
\end{equation}

\noindent where $N$ is the look-back period used to calculate $m_{t-1}(N)$, and $T$ is the total length of our data series (for instance the total number of weeks). By using Eq. (\ref{eq:strategy}), one can rewrite  Eq. (\ref{eq:R}) as

\begin{eqnarray} 
\left \langle R \right \rangle &=& \frac{1}{T-N+1}
\frac{1}{N}\sum_{t=N}^{T}\sum_{i=t-N}^{t-1}X_i X_t \\ \nonumber &=&
\frac{1}{T-N+1} \frac{1}{N}\left[ \sum_{t=N}^{T} X_t X_{t-N} + \sum_{t=N}^{T} X_t
X_{t-N+1} + \cdots +\sum_{t=N}^{T} X_t X_{t-1} \right]\\ \nonumber &=&\frac{1}{N}
\sum_{i=1}^{N} \left \langle X_{t} X_{t-i} \right \rangle 
\end{eqnarray}

\noindent where $<\ >$ stands for average (sometimes represented by $E[\  \ ]$). The last equality is only true if the process for $X$ is such that the product $X_T X_{T-\tau}$ is equal in probability to $X_{T-1}X_{T-1-\tau}$, hence it depends only on $\tau$.

It was pointed in section \ref{sec:stationaryProcess} that in order to exactly model non-stationary data, we need to know how the different stationary patches relate to each other, because the moving average $m(N)$ crosses different patches. Due to this complexity we restrict our theoretical analysis to stationary patches before working with the non-stationary data.

For stationary random variable $X_t$, the expected return (Eq. (\ref{eq:R})) can be expressed as an average of auto-covariance as follows:

\begin{equation}\label{eq:theAve} 
\left \langle R \right \rangle = \frac{1}{N}
\sum_{i=1}^{N} \left \langle X_{t} X_{t-i} \right \rangle = \mu^{2} + \frac{V}{N}
\sum_{i=1}^{N} \rho(t,t-i) 
\end{equation}

\noindent where $\rho$ is the autocorrelation function, $V$ the variance and $\mu$ the mean of the stationary stochastic process $X$. Notice that the result (Eq. (\ref{eq:theAve})) is independent of functional form of the distribution of $X$.

The variance $Var(R)$ of the trading rule in Eq. (\ref{eq:strategy}) is given by:

\begin{equation} \label{eq:risk1} 
Var(R) = \left \langle \left(\frac{X_t}{N}
\sum_{i=1}^{N} X_{t-i}\right)^{2} \right \rangle- \left \langle
\frac{X_t}{N}\sum_{i=1}^{N} X_{t-i} \right \rangle^{2}. 
\end{equation}

The first term in Eq. (\ref{eq:risk1}) relates to the autocorrelation of the squared return and the cross-correlation with the squared return (similar to the leverage effect). The first term can be re-written as:

\begin{equation} \label{eq:risk2} 
\frac{1}{N^2}\left(\sum_{i=1}^{N} \left \langle X_t^{2}
X_{t-i}^{2}\right \rangle + \sum_{i,j=1,i \neq j}^{N} \left \langle X_t^{2}
X_{t-i}X_{t-j}\right \rangle\right). 
\end{equation}

We further simplify Eq. (\ref{eq:risk1}) by assuming a multivariate Gaussian distribution. Thus, the correlations are linear correlations and the marginal distributions are Gaussian. Although empirical financial data is not described by a Gaussian distribution we will see that the weekly normalized returns obtained according to Eq. (\ref{eq:normalizeData}) in section \ref{sec:E_stationary} are well approximated by a Gaussian distribution.

We can, therefore, calculate the variance of our trading rule by using the characteristic function of the multivariate Gaussian distribution. Performing the right order of differentiation and enforcing that the variance $V$ and the drift $\mu$ of the returns are constants and   that the autocorrelation depends only on time lags, we have the variance of our trading  rule given by:

\begin{eqnarray}\label{eq:theVar} 
Var(R) &=& \frac{1}{N^{2}}\bigg[ N V^{2} + N \mu^{2} V
+ N^{2} V \mu^{2} \\ \nonumber &+&V^{2} \big(\sum_{i=1}^{N} \rho(t,t-i)\big)^{2} +V^{2}\sum_{i,j=1,i\neq
j}^{N} \rho(t-i,t-j)\\
\nonumber &+&\mu^{2} V \Big(2\sum_{i=1}^{N} \rho(t,t-i) +\sum_{i,j=1,i\neq
j}^{N}\big(\rho(t,t-j)+\rho(t-i,t-j)+\rho(t,t-i)\big)\Big) \bigg]  
\end{eqnarray}

\noindent where $\rho(t,t-i)$ is the correlation coefficient of the returns from time $t$ and $t-i$. Details of the calculation can be found in appendix \ref{sec:app1}. In the next section we will derive useful asymptotic limits of Eq. (\ref{eq:theVar}).

\subsection{Limits and interpretations}

In contrast to prior studies \cite{Kim,Lewellen,Lo1990,Moskowitz2011}, we do not look only at the return of the trading rule. We use the variance (\ref{eq:theVar}) to calculate the Sharpe ratio (SR), defined here by the ratio between the average return and     the standard deviation (square root of the variance).\footnote{1It is usual to define the Sharpe ratio by removing the interest rate from the mean return. However, note that we have assumed that the interest rate is zero. It is also worth noting that the Sharpe ratio (SR) is a special case of the Information ratio (IR).   The IR measures performance against a benchmark which might be in general a risky portfolio. However, for the purpose of this study our benchmark is cash. Therefore, SR is equal to IR for   us.} The general expression is fairly complex but two limiting cases are enough to help us understand how the SR depends on the parameters.

\begin{figure}[htbp] \centering
\includegraphics[width=\linewidth]{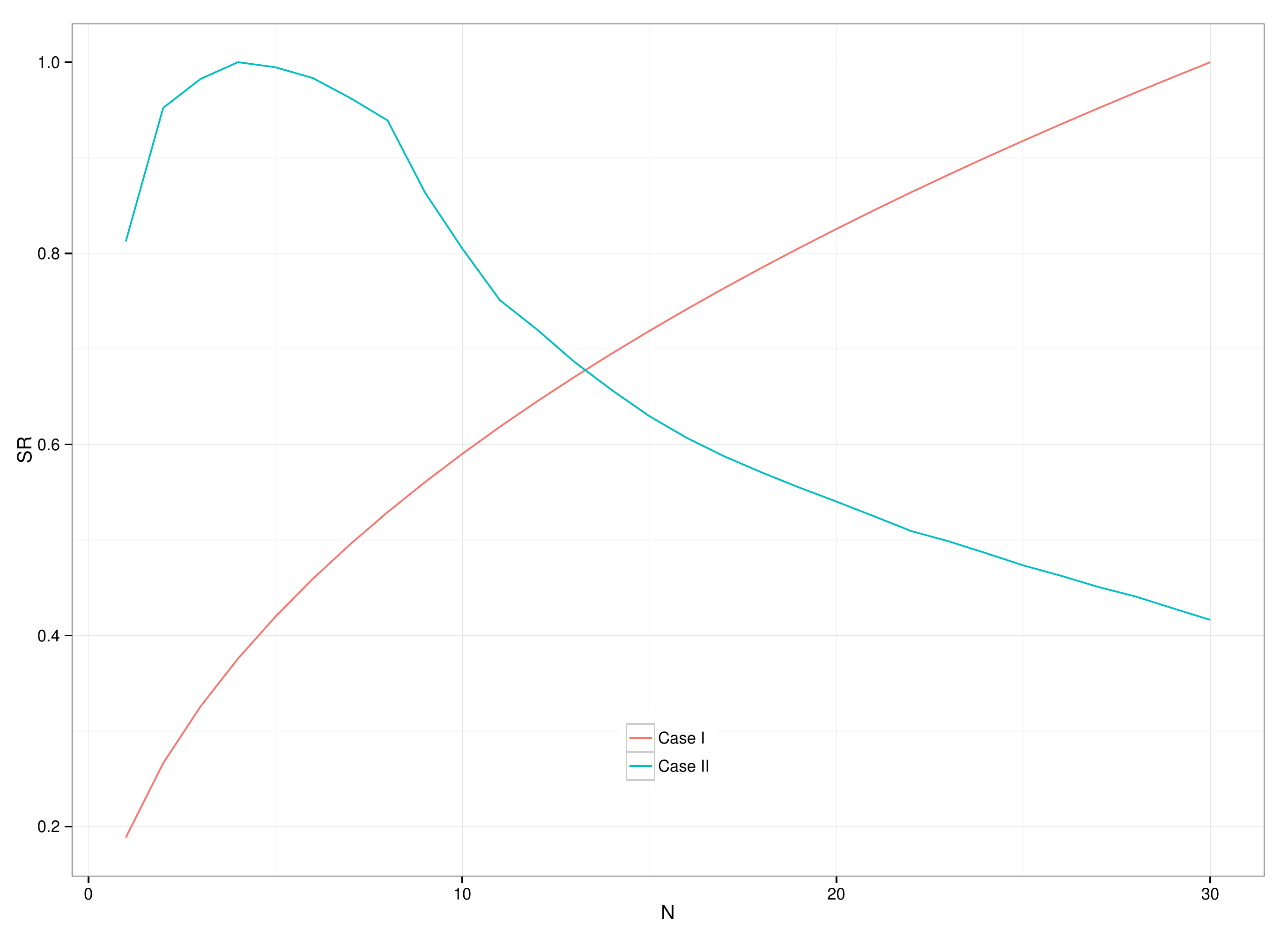} \caption{Increasing red line: SR as a function of the look-back lag $N$ for case I ($\rho=0$, Eq. (\ref{eq:case1})) normalized to have maximum of 1. Decreasing blue line: Normalized SR for case II ($\mu=0$, Eq. (\ref{eq:case2})) generated using simulated data
from an hypothetical process with $\rho \neq 0$ for lag 1 to lag 5.\label{fig2}} 
\end{figure}

In case I, all the autocorrelations are zero: $\rho(t,t-i)=0$. This is equivalent to say that the log-returns are independent and identically distributed (IID) Gaussian random variables. The Sharpe ratio (SR) is given by:

\begin{equation}\label{eq:case1} 
\text{SR} =
\frac{\mu^{2}}{\sqrt{V\mu^{2}+\frac{V^{2}}{N}+\frac{\mu^{2}V}{N}}} 
\ \xrightarrow[N \to \infty]{} \
\frac{\left|\mu\right|}{\sqrt{V}}, 
\end{equation} 

\noindent where $N \rightarrow \infty$ is the limit of long (or short) and hold since $m(N)$ converges to $\mu$. It is interesting that the optimal point is when $N \rightarrow \infty$. In this case, the best Sharpe ratio is in fact the one that would
be expect for a given process: mean over standard deviation. Any other $N$  gives worse results. Hence, as expected, if the given process is iid, the moving average $m(N)$  provides one way to estimate $\mu$.   A representation of the SR as a function of $N$ for case I is given in Figure \ref{fig2}.
In case II, we assume that $\mu=0$. Thus, all performance comes from autocorrelation. The SR is given by

\begin{equation}\label{eq:case2} \text{SR} =
\frac{\sum_{i=1}^{N}\rho(t,t-i)}{\sqrt{N+(\sum_{i=1}^{N}\rho(t,t-i))^{2}+(\sum_{i,j=1,i\neq
j}^{N}\rho(t-j,t-i))}} \end{equation}

\noindent where the exact shape of SR, as a function of $N$, depends on the way $\rho$ varies as a function of $N$. Practically it is very unlikely that $\sum_{i=1}^{N} \rho(t,t-i)$ grows fast enough to dominate the $\sqrt{N}$ term in the denominator. More surprising is that Eq. (\ref{eq:case2}) does not depend on the variance $V$,  in other words, the SR of the trading rule is the same for very large or small $V$. The expression is also of practical use      since one can calculate SR for a given correlation value. For instance, for $N=2$ weeks, $\rho(t,t-1) = 0.05$ and $\rho(t,t-2) = 0.02$, SR $ \approx 0.0422$ per week.

In order to illustrate case II, we look at a moving average process ($MA$ i.e. of the kind $x(t) = a_{1} \epsilon_{1}(t)+a_{2} \epsilon_{2}(t-1)+..., \epsilon_{i} \sim N(0,\sigma)$)  with  autocorrelation  $\rho \neq 0$  from  lag  1  to lag  5.  The general shape of case II is presented in Figure 1 (case II, blue line). We have chosen to show an arbitrary $MA$ process because it is the only process from the $ARMA$  family where the $\rho$ term in the numerator can grow fast enough before $1/\sqrt{N}$ takes over. This creates the “hump” in the graph. The hump is located where the $MA$ process has autocorrelation different    from zero and the size of the hump depends on the intensity of the autocorrelation. 

In summary, case I (red line) is increasing with the value of $N$ , while case II (blue line) increases initially and then decreases slowly (\ref{fig2}).

It is clear that if we have a pure case II, there could be an optimal $N$ above which we would get worse risk adjusted performance. Therefore, it is generally not advantageous to use the trading rule (\ref{eq:strategy}) unless the best $N$ is selected.

The dependence of the SR on N is different for case I and case II: the first case grows and the second decreases in the limit of a large N. Normally, empirical data has SR as a hybrid of case I     and case II, and that indicates the trading rule (\ref{eq:strategy}) will transit from a case II dominated performance to a case I dominated performance with the increase of $N$ . Large $N$ shows that case I is targeted: the moving average estimates the drift. Keep in mind that if the sum of the autocorrelations happens to be positive (negative), case I is shifted up (down) due to the case II contribution.

Finally, we remind the readers that our discussion here assumes a stationary process. That is clearly not the situation in practice. We assume that there are patches or periods where the data is approximately stationary. Next, we will look at data by approximately finding these stationary patches before dealing with non-stationary data

\subsection{Computer simulated data analysis}

The computer simulation approach allows to track the evolution of complex systems, to investigate  aggregate behaviour and
to look for emergent phenomena. This approach has been largely applied
in many areas, in particular to study the financial markets, where simulated time series may exhibit the main empirical properties of real financial markets, called stylized facts \cite{cont2001empirical}. Several frameworks were explored in the literature, for instance, the minority game model \cite{challet2001stylized, crepaldi2009multifractal}, the stochastic multi-agent model \cite{lux1999scaling} and Arch/Garch models \cite{engle2001good}.

One can also use autoregressive moving average processes (ARMA) to simulate stationary or nonstationary time series \cite{montgomery2015introduction}.  This kind of process combines two 
types of dependence of present observation $z_t$ with past events. In other words, current observations depend on the level of its lagged observations $z_{t-p}$ 
(autoregressive process of order $p$, denoted by AR$(p)$) and also depend on observations of a random variable at present time  $\epsilon_t$ as well as the shocks that have taken 
place before $ \epsilon_{t-q}$, (moving average process of order $q$, denoted by MA$(q)$). The mathematical expression to represent the ARMA$(p,q)$ is:
\begin{equation}
 z_t=\phi_0+ \phi_1 z_{t-1} + \phi_2 z_{t-2}+...+\phi_p z_{t-p}+\epsilon_t+\theta_1 \epsilon_{t-1} + \theta_1 \epsilon_{t-2}+...+\theta_q \epsilon_{t-q}
\end{equation}
where, $\phi$ and $\theta$ are constants and shocks are normally distributed, i.e., $\epsilon \sim N(0, \sigma_{\epsilon}^2)$. 

It is not difficult to obtain non-stationary time series which originally may be due to trends mostly found in the mean or the instability in the variance or both. For instance,  the ARMA model with time-dependent coefficients may easily exhibit non-stationarity. In general, the ARMA model is not guarantee to generate stationary time series. Therefore, one can use some statistical test to check for stationarity, such as Augmented Dickey-Fuller test \cite{said1984testing}, KPSS-test \cite{kwiatkowski1992testing} or Leybourne-McCabe stationary test \cite{leybourne1994consistent}.

In this section, we validate the theoretical calculations developed in the previous section by using simulated stationary ARMA$(p,q)$ log-returns. Although we will test our theoretical results against real data in the following sections, here we want to investigate how good is the agreement of the Sharpe ratio defined by Eq. (\ref{eq:theVar}) and Eq. (\ref{eq:R}) versus the Sharpe ratio of simulated ARMA$(p,q)$ stationary log-returns with Gaussian noise. 
 
\begin{figure}[htbp] \centering
 \includegraphics[width=\linewidth,height=12cm]{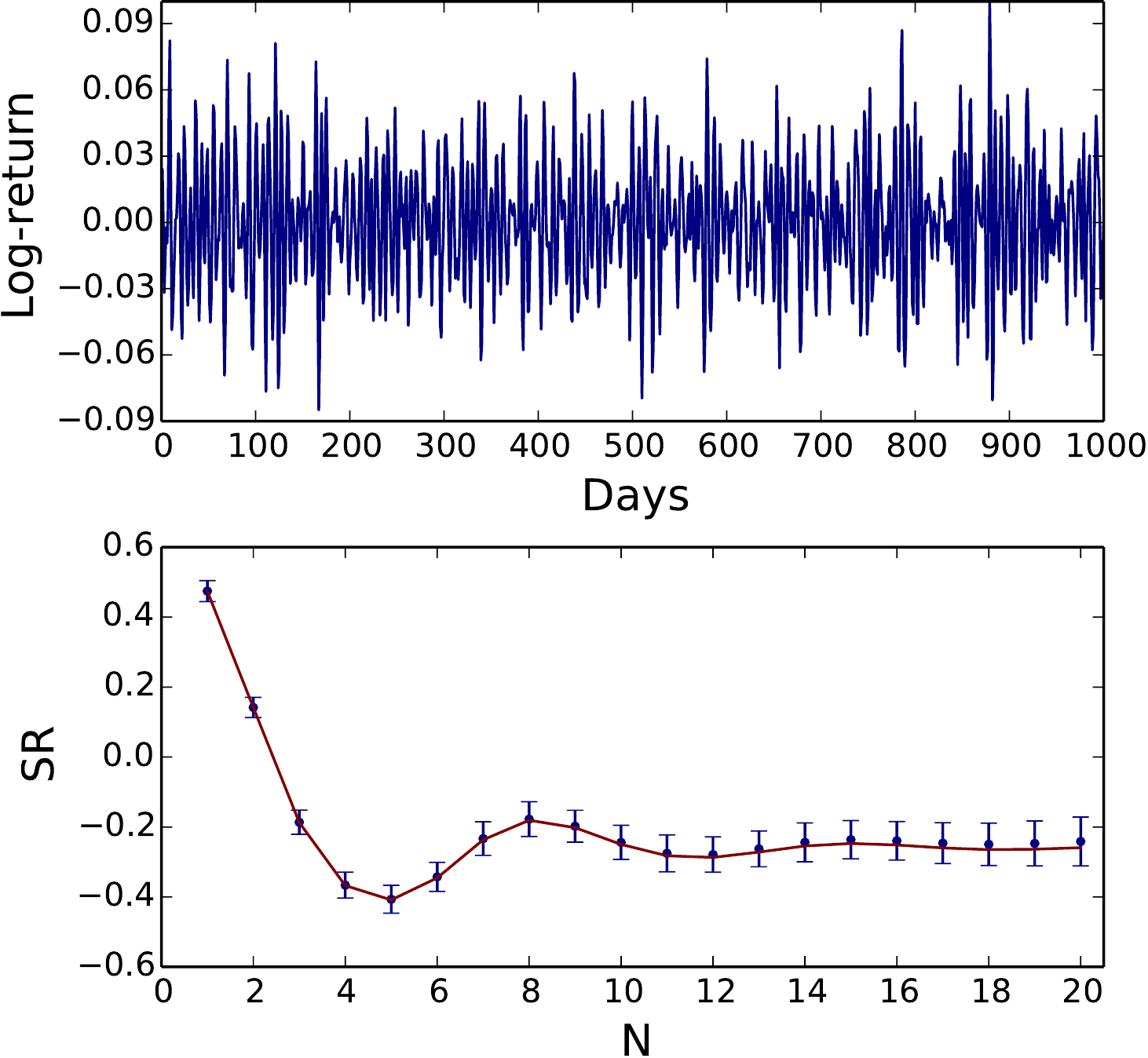} \caption{ In the upper panel: Time series of Log-return simulated from an ARMA(2,2) with the following parameters: $\phi_0=0.9$, $\phi_1=0.95$, $\phi_2=-0.6$, $\theta_1=1.4$, $\theta_2=0.5$ and variance $\sigma_{\epsilon}^2=0.3$ (volatility of approximatly 0.55 per year or 0.035 per day). In the lower panel: Sharpe ratio from simulated data using the same time series ARMA(2,2). Simulated SR are averaged over 200 realizations (solid circles), with error bars indicating one standard error. The continuous line is the predicted Sharpe ratio using Eq. (\ref{eq:R}) and the square root of Eq. (\ref{eq:theVar})  \label{fig:SRS}} \end{figure}

We perform our computational experiments in the following way. First, we define the ARMA parameters and generate one sample with a similar volatility of real financial data and size equals to 2000 points (days). Then, we check stationarity using Augmented Dickey-Fuller, KPSS and Leybourne-McCabe tests. If stationarity is confirmed, we generate 200 time series with the same set of parameters. All statistics are computed by considering all realizations. In Figure \ref{fig:SRS} we show, in the upper panel, one single realization of a stationary ARMA(2,2) (we depict only 1000 points for simplicity). In the lower panel of Figure \ref{fig:SRS} we show the average of SR as a function of $N$, which is a result of applying strategy Eq. (\ref{eq:strategy}) to every ARMA(2,2) simulated time series. Error bars were obtained by using the standard deviation of the Sharpe ratio over all time series (the standard error, defined here as the standard deviation divided by the square root of the number of realizations). The solid line is the theoretical prediction for Sharpe ratio using Eq. (\ref{eq:R}) and the square root of Eq. (\ref{eq:theVar}). We repeat the experiment for different parameters of ARMA$(p,q)$ model, and for all, we find a good agreement of theoretical and simulated data.  
  
\section{Empirical analysis} 

We use the Dow-Jones Industrial Average (DJIA) Index from 05/1896 to 07/2015 to perform our empirical analysis. The daily DJIA index values are downloaded from Federal Reserve Economic Data (FRED) webpage.\footnote{Data download uses ``quantmod" library in R \cite{R}.} We convert the daily index values to weekly index values and calculate weekly log-returns. We choose to work with weekly data because we get at least four times more data than the traditional monthly values and also reduce considerably any market micro-structure issues from daily data.

We present this study with the DJIA index but we  also looked at several other American and international  indices.  Since most  of  the  indices  do  not  have  the  same  breath  of  data  as  the  DJIA  we have chosen to present results  based  on  the  DJIA.  Appendix  \ref{sec:app3}  shows our  main  result  (SR  as  a function  of N) applied to other indices with  virtually  the  same  qualitative results.

Most prior studies have  a minimum holding period of one month even when using weekly data     \cite{Kim}. In contrast with them we  re-balance our position weekly (holding period         of one  week),  which  maximizes  data  usage.  We use the  software  {\bf R} \cite{R} for our analysis. A simple sample code pertinent to this article can be found in \cite{Rpubs}.

In appendix \ref{sec:app4} we apply our method using both monthly data and holding period of one month as well        as daily data with holding period of one day. The main result holds, irrespective to the data/trading frequency selected.

Weekly index levels/prices are not uniquely defined. There are five different weekly return data sets:  Monday to  Monday,  Tuesday  to  Tuesday  up  to  Friday  to  Friday.  We perform  our  analysis  on all five definitions. Furthermore, when calculating weekly returns we make sure to filter out     time  gaps  larger  than  one  week  (seven  days),  that  is,  we  remove  from  the  data  special  situations (such as the 1914 stock market trade halt due to the First World War) and holidays. In conclusion, our empirical work is performed on five different time series but when presenting the results in this section we average over all those  series.

\begin{figure}[htbp] \centering
\includegraphics[width=\linewidth]{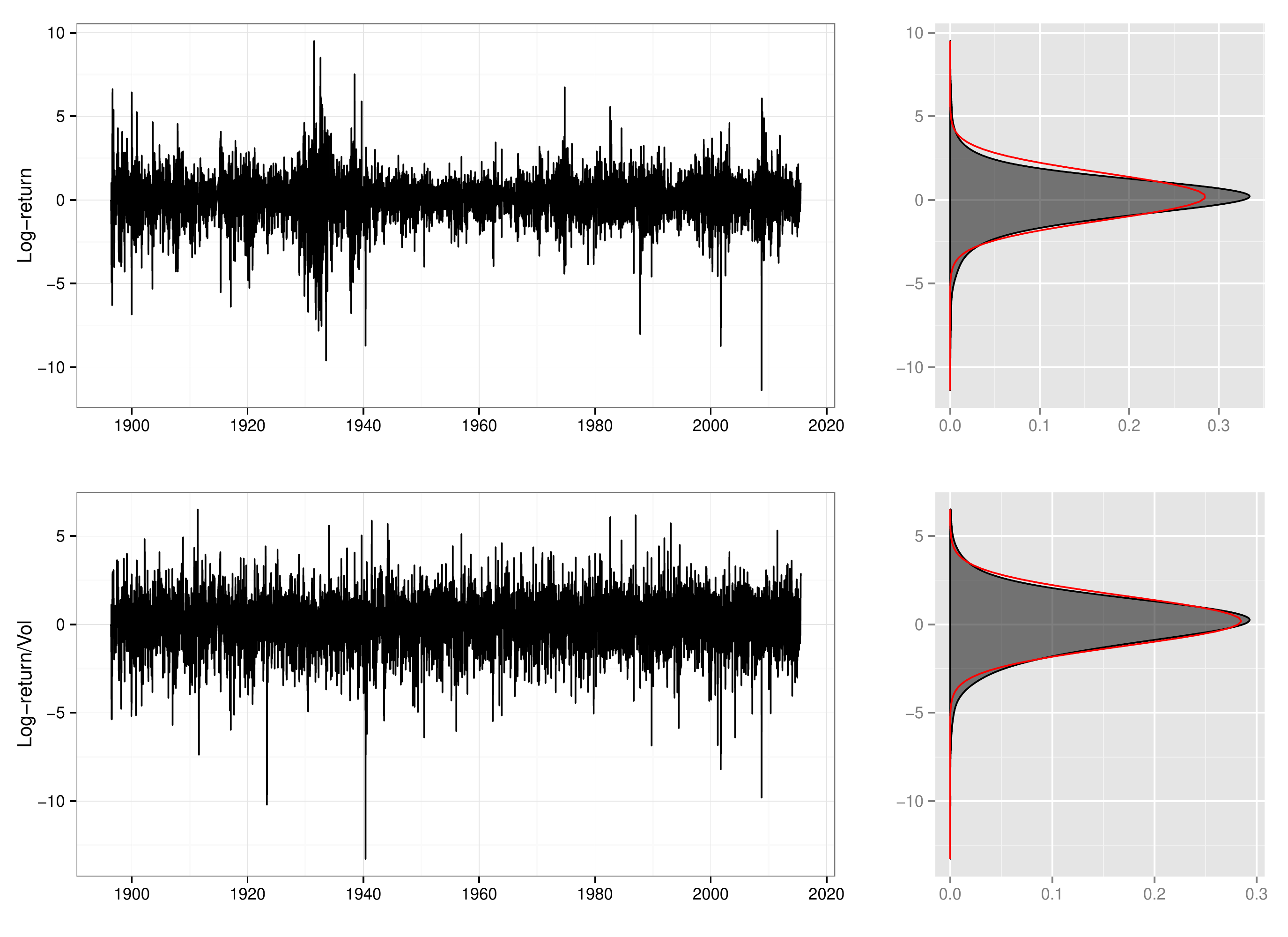}
\caption{Time-series (left) and corresponding probability distribution functions (PDF) [right, black solid line] for   the log-returns [top two graphs] and normalized log-returns using (\ref{eq:normalizeData}) [bottom two graphs]. In order for the log-returns (top) to show a comparable y-axis we have normalized the returns by the in-sample average absolute log-return of the whole history. Therefore, the values of the returns are unrealistically large (easily around one). Notice that the normalized log-returns show a much more stable volatility and consequently the PDF (black solid line) is closer to a Gaussian (red solid  line). 
\label{fig:volClu}} 
\end{figure}

The model in section  \ref{sec:t_model} uses log-returns (Eq. \ref{eq:logR}), however, an investor is paid according to linear returns ($S_{i}/S_{i-1}-1$). Therefore, if the analysis wants to have a practical application, it is necessary that $ln(S_{i}/S_{i-1}) \approx S_{i}/S_{i-1}-1$. We have verified that this is indeed the case for weekly returns of the DJIA index. Furthermore the equivalence of linear  and  log-returns  is  a  common  assumption  for  short  return horizon but it can result into problems if used blindly \cite{Meucci}.

In order to implement the algorithm (Eq. (\ref{eq:strategy})) described above, we first reduce the effect of volatility clustering. We  do so by dividing the returns by the average absolute returns of the past  $p$  periods (Eq. (\ref{eq:normalizeData})). This transformation avoids  look-ahead bias and creates a strategy that can      be implemented realistically.

\begin{equation}\label{eq:normalizeData} 
X_t = \frac{\ln (S_{t}/S_{t-1})}{\sum_{i=1}^{p} \mid \ln (S_{t-i}/S_{t-1-i})/p \mid} 
\end{equation}

We apply  this  normalization  for  both  in-sample/stationary  and  out-of-sample/non-stationary data analysis. This simple normalization reduces volatility clustering and also helps to bring the returns closer to a Gaussian distribution. Figure \ref{fig:volClu} shows that the volatility clustering is greatly reduced and that the final probability distribution is closer to a Gaussian at the center of the distribution. There are still substantial tail events which effects we ignore for this study.

The effect of the normalization is shown in appendix \ref{sec:app2}. Again, the main qualitative results are          the  same  whether  we  use  or  not  the  normalization  (Eq.  \ref{eq:normalizeData}).  That  shows  that  the  results  here  are not an artifact of the    normalization.

\subsection{In-sample and stationary}
\label{sec:E_stationary}

In this section we take the view that financial log-returns are generally not stationary \cite{Guidolin,McCauley,Seemann, fff}. Notice that this assumption is different from several prior studies, e.g. \cite{Bondt1989,Lehmann,Lo1988,Lo1990} which consider 25 years of data (1962 to 1986). This section is more in line with studies that advocate business cycles dependence, such as \cite{Chordia}. Therefore we will assume that both the drift and the autocorrelation are not constant for the entire history. We first deal with the changing drift by breaking the data up into periods where the drift is approximately constant. Next, we show that the autocorrelation does     not have the same sign throughout history. In fact, there is a major change of sign in 1975.

\subsubsection{Dealing with changing drift}

We use the “Breaks for Additive Season and Trend”” (BFAST) algorithm to find the constant drift periods. Here we are interested in the trend component which is a piecewise linear function where the breakpoints (change from a linear function to the next) are determined by the BFAST as well. We apply BFAST to the log DJIA index sampled monthly. In contrast to weekly returns, we define the monthly index levels uniquely as the index value at the last trading day of the month. We choose a monthly index because running the algorithm          is faster and because we  want  periods  that  are  few  years  long  in  order  to  have  enough  data points (weeks) within each regime. Our average regime length is of 2.2 years with a maximum        of approximately five years, and a minimum of approximately one year.

\begin{figure}[htbp] \centering
\includegraphics[width=\linewidth]{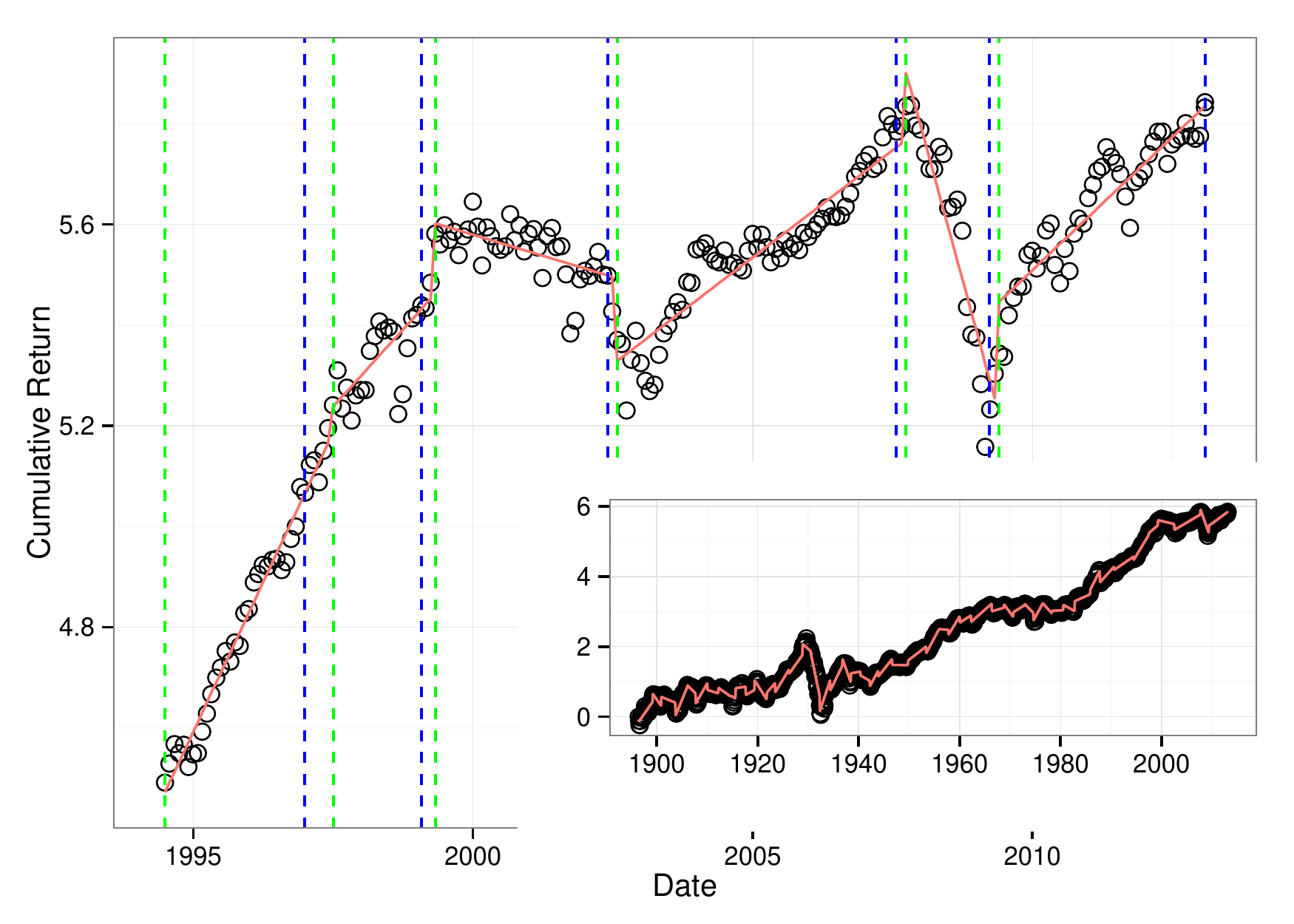} \caption{Log cumulative return of weekly DJIA from 1995 to 2013 (circles). The intervals between green and red vertical dash lines were obtained by using the BFAST algorithm which finds these intervals by fitting piecewise linear functions to the data. The linear fits are represented by the solid red lines. \label{fig4}} \end{figure}

After applying BFAST to find the stationary patches, we ignore any regime with less than 1.3 years (70 weeks) in order to have enough data to apply our trading rule with   $N$ close to 1 year.      We are left with 47 regimes (patches). We plot the log cumulative return for DJIA from 1995 to   2013 together with the fitted trends in Figure \ref{fig4}. Each regime is defined between dashed green line and the consecutive dashed blue line. The inset shows the data from 1900 to 2013.

In general, the linear regression in Figure \ref{fig4} shows a very good fit. Some patches present a good adherence with the linear fit while others present a larger fluctuation around the linear fit. The goodness of the fit validates the method used  in  this  work.  Patches with  larger  fluctuations  correspond  to troubled economical moments as in 1997, 2002 and  2008.

\subsubsection{The autocorrelation is not stationary}

\begin{figure}[ht] 
\centering \includegraphics[height=8cm,
	width=\linewidth]{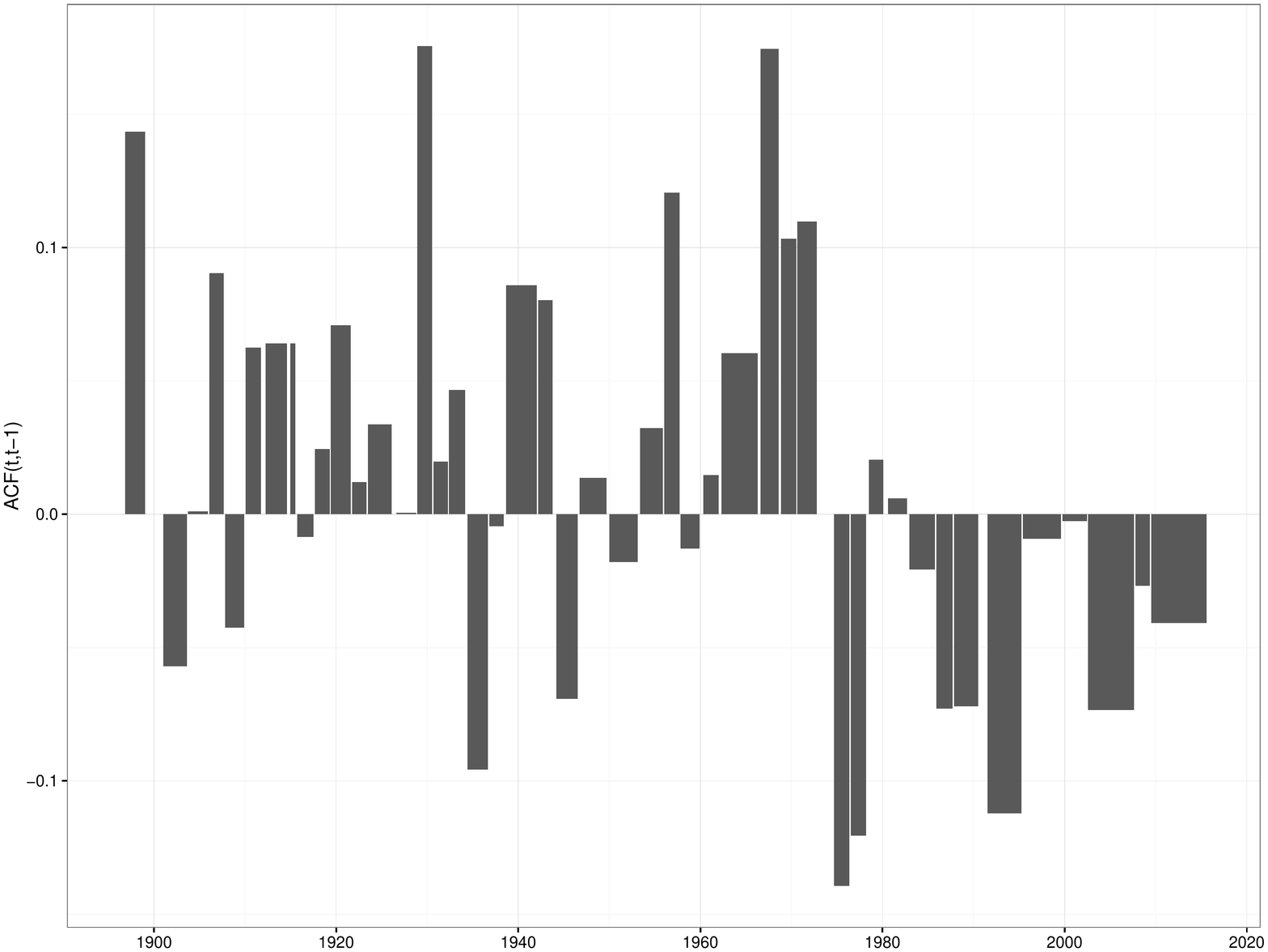}
\caption{First lag autocorrelation for rescaled DJIA log-returns within each regime. Notice the clear change in sign around 1975 which lead us to conclude that the autocorrelation is not stationary through the nearly 120 years of data.\label{fig6}} 
\end{figure}

In order to understand the stability of the autocorrelation across regimes/patches, we calculate the first lag of the autocorrelation for the weekly log- returns within each regime (Figure Figure \ref{fig6}). Notice  that  from  1896  to  1975,  most  autocorrelations  are positive, after 1975 the autocorrelations are mostly negative. The sign of the first lag autocorrelation survives even if we calculate the autocorrelation over all 4080 weeks up to 1975 and all 2001   weeks
after 1975. The average of all weeks (Monday-Monday, etc) first lag autocorrelation returns before     1975 is 0.05 and after 1975 it becomes $-0.04$  (both significant at 95\%). Thus, there is a change to the sign of the autocorrelation of weekly returns around 1975.

Previous work by \cite{Lo1988,Lo1990} hinted that the autocorrelation regime changes reported here (Fig. \ref{fig6}). They showed that for an index of large cap stocks, the autocorrelation for
the sub-period (1975$-$1985) is not significantly positive, although being significantly positive   from
1962 to 1988. See also the work of \cite{Froot}, that studied the first lag autocorrelation
for returns from 15 minutes to one week for stock indices with data up to 1990. In agreement with our results they show that the large positive autocorrelation decreases substantially after 1970,   going negative (see Figure 5 in \cite{Froot}).

To  further compare our results to  \cite{Lo1988,Lo1990}, we calculate the first four     lag autocorrelations of the weekly log-returns, from July 6, 1962 to December 31, 1987, for the DJIA index. Our values have the same signs as reported by  \cite{Lo1988,Lo1990} but different magnitudes. We get 1.1\%, 1.7\%, 5.1\%, $-$2.8\% for lags 1 to 4 whereas they had 7.4\%, 0.7\%, 2.1\% and $-$0.5\%  (for the  value-weighted CRSP  index, see Table  1 of  \cite{Lo1990}).  We
attribute these different values to the fact that we use the DJIA and they looked at the value- weighted Center for Research in security Prices (CRSP) index.

The autocorrelation values shown in Figure \ref{fig6} are calculated with the DJIA rescaled log-returns (Eq. (\ref{eq:normalizeData})). Therefore we expect an additional difference, especially on the magnitude of the autocorrelation. We find that the first four autocorrelation of our rescaled DJIA index are 5.7\%,1.6\%,0.7\% and -3.2\% which agree in sign with both Lo and MacKinlay (7.4\%, 0.7\%, 2.1\% and $-$0.5\%)      and with the un-scaled DJIA log-returns (1.1\%, 1.7\%, 5.1\%, $-$2.8\%). The positive sign of the first autocorrelation only shows that by using data from (1962$-$1988) we are missing the regime change around 1975.  In other words, by taking the data from 1962 to 1988, we are mixing our 11 regimes   in such a way that the first lag autocorrelation becomes positive, we find that the pre-1975 data “dominates”     the post-1975 data when the autocorrelation of the entire period (1962 to 1988) is calculated.

For completeness, it is worth mentioning that \cite{Bondt1989,Lehmann} report  that individual weekly stock  returns  are  negatively  autocorrelated  using  data  from  virtually  the  same period (1962  to  1986)  as  in  \cite{Lo1990}.  This  apparent  dilemma  (since  indices  are made of individual stocks) was addressed  by  \cite{Lo1990}.  They  show  that  it  is possible for the autocorrelation of the index to be positive even though the autocorrelation of the component stocks is negative if the stocks cross-autocorrelation (lead-lag relation among stocks) is   large and positive.

Our  autocorrelation  results  in  Figure  \ref{fig6}  indicate  a  fundamental  change  that  starts  in  1975  for  at least  large  cap  stocks.  Not  only  is  the  first  lag  autocorrelation  of  DJIA  weekly  returns  negative but  assuming  that  first  lag  autocorrelation  for  individual  stocks  is  still  negative  (see  \cite{Gutierrez} with data from 1983  to  2003),  cross-autocorrelation  between  stocks  should  be negative or insignificant after  1975.

\subsubsection{Results: SR vs N}

\begin{figure}[htbp] \centering
\includegraphics[width=\linewidth]{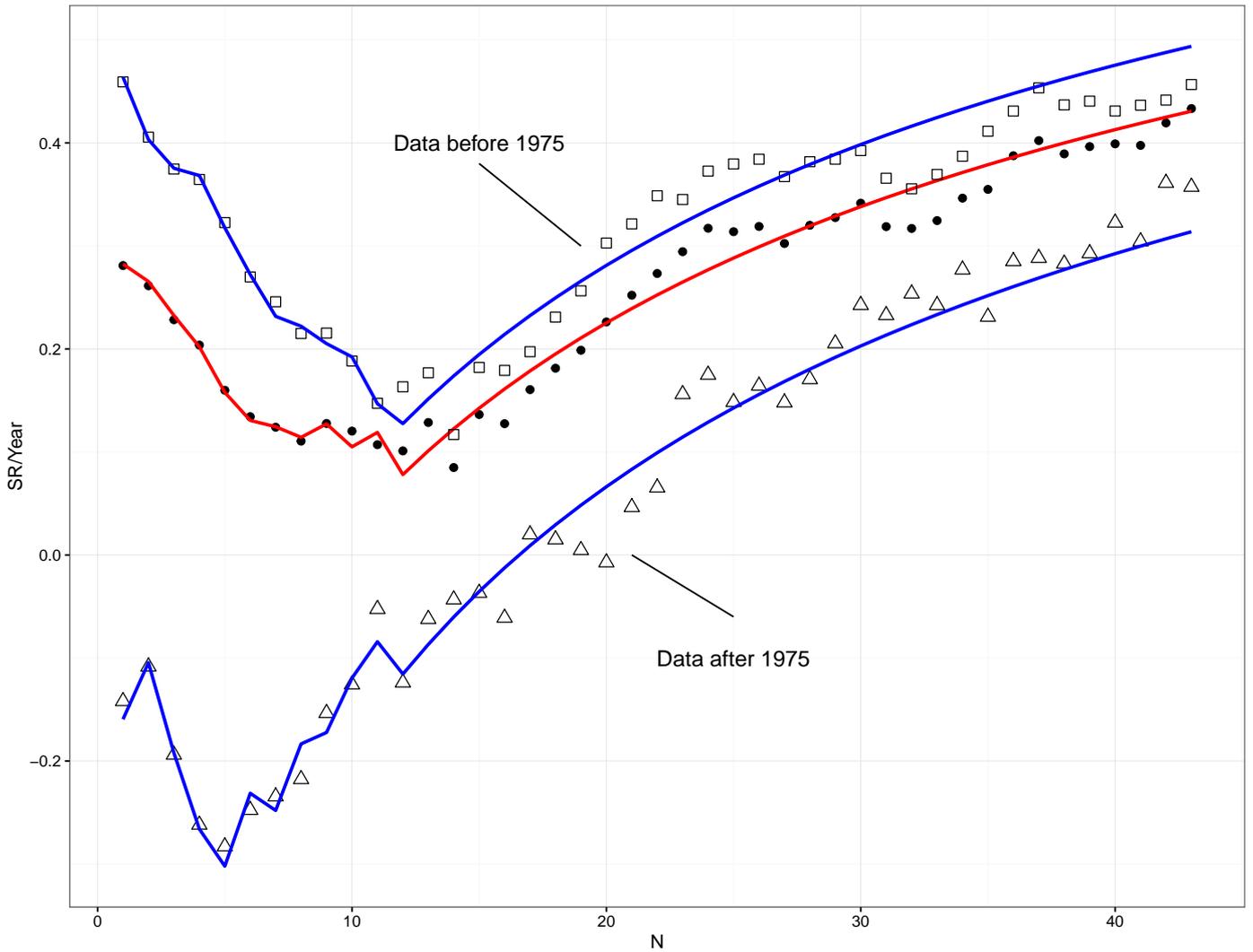}
\caption{Average SR vs. $N$ (look-back)  over patches determined by BFAST algorithm. Average is performed over regimes before 1975 (squares) and after 1975 (triangles) as well as all 47 regimes/patches (circles). Symbols represent the data and the solid lines are the best fit theoretical SR given by Eqs. Eqs. (\ref{eq:theAve}) and
(\ref{eq:theVar}).\label{fig:aveSpec}} 
\end{figure}

We calculate the empirical performance of the trading rule  (Eq.  (\ref{eq:strategy})) for look-back periods ranging from $N=1$ to $N=43$ weeks within each regime. Figure \ref{fig:aveSpec}    shows the average of SR vs N (look-back) over all 47 regimes. In addition   to the ensemble average, we  show  the  theoretical  SR  that  is  constructed  by  dividing  the  theoretical  average  performance (Eq. (\ref{eq:theAve})) by the theoretical  standard  deviation  of  the  performance  (Eq.  (\ref{eq:theVar}))  of  section  \ref{sec:t_model}.  To  be precise, we first compute the average of SR vs N for each week (starting on Monday,  starting on  Tuesday,  etc.),  perform  the  regime  average  as  well  as  theoretical  fit  on  each  week  independently and then average over all five time series (week with different starting day).

The parameters for the theoretical model are found by least squares estimation fitting the empirical data points. The model uses the empirical standard deviation ($1.43$)  of our renormalized data and 13 fitting parameters: the drift $\mu$ and the first 12 autocorrelations.  Clearly we  are  over-fitting,  since  the number of parameters is large giving that we have only a maximum of 43 data points. However,    our goal here is not to find a robust model but to qualitatively show the potential of the theoretical model.

Figure \ref{fig:aveSpec} shows the SR vs $N$ of our strategy for the average of all regimes pre-1975 and post-1975 as well as the average over all regimes. The theoretical formula describes well the data and we      can see that the graph for pre- and post-1975 are very different due to the change from positive        to negative autocorrelation of the returns documented in Figure \ref{fig6}. The overall average (red solid line) is similar to the pre-1975 and considerably different from the post-1975 especially for small $N$. That shows that when performing an average over all regimes the pre-1975 data dominates the post-1975 data.

Independent of the period we look, Figure \ref{fig:aveSpec} confirms the expectation that the data presents a
mixture of case I and case II. For a very short look-back, one or two weeks, we have a large positive or negative autocorrelation effect (case II). For look-back periods of more than $N=16$ weeks ($\approx$ four months) for pre-1975 or $N=5$ weeks ($\approx$ one month)  for post-1975 data, the SR curves in
Figure \ref{fig:aveSpec} clearly resembles case I, where if we  add more look-back weeks, we  do better. In fact, case I is on average stronger (larger SR) for large enough look-back than case II.

In our model, case I (drift) has a positive effect no matter what the sign of the DJIA drift          ($\mu$ in Eq. \ref{eq:case1}) while the effect of case II (autocorrelation) can also be negative depending on the     sign  of  the  return  autocorrelation  ($\rho$ in Eq. \ref{eq:case2}).  Therefore,  in  Figure  \ref{fig:aveSpec}  the  relative  importance  of case II may be underestimated if the autocorrelation does not have the same sign throughout all the regimes.  However the  most  important  factor  that  makes  drift  as  significant  as  the  autocorrelation (for data pre-1975) is that our patches/regimes  are  defined  to  have  a  well  specified  drift.  That  is optimal for the strategy and therefore the average over the regimes is an average over the absolute   value  of the drift (Eq. \ref{eq:case1}), and not over  the drift which is clearly smaller. In Section \ref{sec:NE_stationary} we  remove       the regimes to show that the effect of the drift cannot be easily recognized from the graph of SR vs $N$. We construct SR vs $N$ for all 119 years of the DJIA. The goal is to capture out-of-sample/non- stationary large $N$  effects \cite{Bondt1989} and to simulate the way  a strategy is typically tested        in  applications.  We show  that  our  stationary  SR  formula  does  not  apply  unless  one  is  to  assume a very complex and long-ranged autocorrelation, so we  propose a simple non-stationary model for      the index return which describes SR across N   well.

\subsection{Empirical non-stationary analysis}
\label{sec:NE_stationary}

In this  section  we  apply  the  moving  average  strategy  of  Eq.  \ref{eq:strategy}  to  the entire  DJIA data.  We  do  not  try to  break  the  data  into  stationary  periods,  but  we  still  normalize  the  weekly  log-returns  using  Eq. \ref{eq:normalizeData}. The effect of normalizing the returns is discussed in appendix \ref{sec:app2}.

The goal here is to present the effect of non-stationary data to the performance of moving average strategies. Since we normalize the log-returns using Eq.  \ref{eq:normalizeData}  we  still  expect  the  data  to  have constant variance. Nevertheless, the data will not have constant drift nor constant autocorrelation. Therefore, we do not expect the results of this section to conform with case I or case II in Figure \ref{fig2}.

\begin{figure}[htbp] \centering
\includegraphics[width=\linewidth]{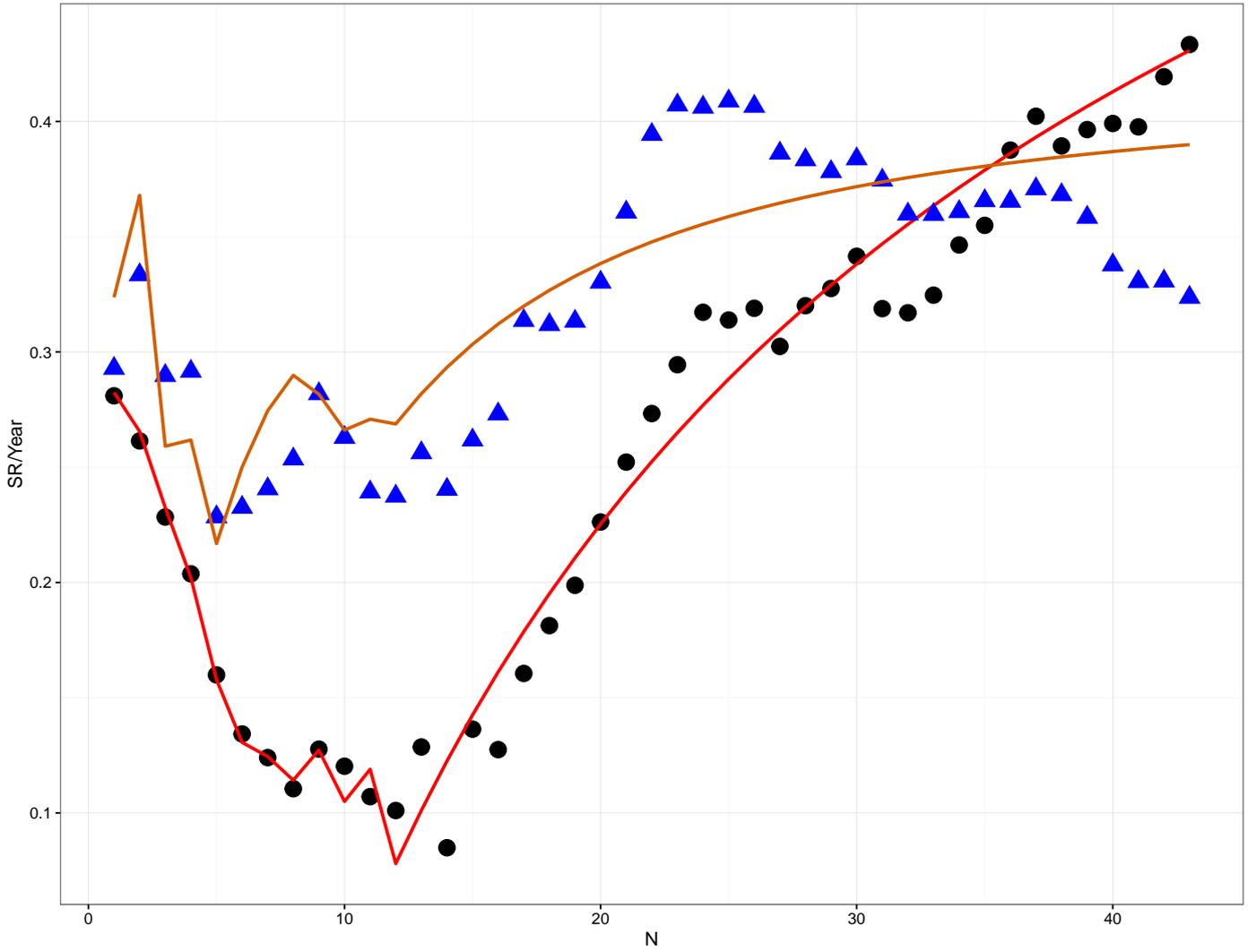} \caption{Average SR vs. N (look-back) over patches determined by BFAST algorithm. Average is performed over regimes before 1975 (squares) and after 1975 (triangles) as well as all 47 regimes/patches (circles). Symbols represent the data and the solid lines are the best fit theoretical SR given by Eqs. (\ref{eq:theAve}) and (\ref{eq:theVar}).
\label{fig:SRcomp}} 
\end{figure}

\begin{figure}[htbp] \centering
\includegraphics[width=\linewidth]{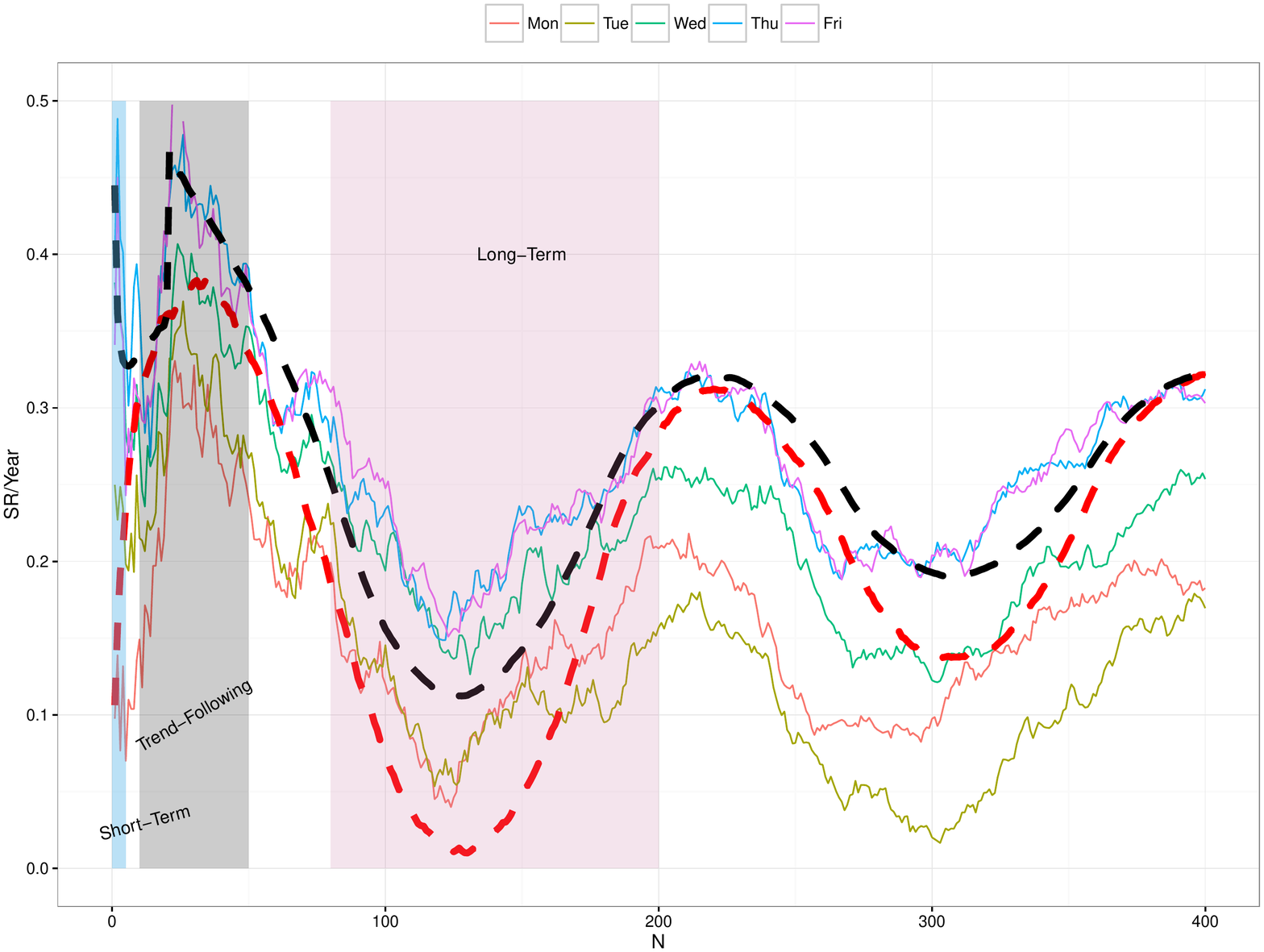}
\caption{SR vs. portfolio formation period $N$ in weeks. The thin solid lines represent the empirical SR per trading day. The highlighted areas point to the range of $N$ reported in the literature (mostly stocks). Our study goes beyond the range normally described in the literature and uncovers a periodic oscillation on the SR. Dashed lines is the result of theoretical model from Eq. \ref{eq:nonSmodel}. The lower red line is generated with IID Gaussian distribution and higher dashed blue line is generated with an auto-correlated Gaussian distribution. The 119 years of data imply that SR values larger
than $\approx 0.2$ are significant at  95\% according to \cite{bouchaud2003theory}.
\label{fig:aveSpec100}} 
\end{figure}

Figure \ref{fig:SRcomp} shows the results of the previous section (stationary/in-sample average over all regimes
by black circles and corresponding fit) and the SR calculated by applying the strategy over all 119 years of DJIA (non-stationary/out-of-sample: blue triangles and corresponding fit) for the same range of $N=1$ to $N=43$. The most striking difference is that there is no clear drift phase (case
I) when compared to the stationary data. The case I, if present, is reduced to a limited range between $N=15$ and $N = 25$ and quickly reverts when the SR gets progressively worse. Therefore     the assumption of constant drift is only an approximation if we apply the trading rule over all the data.

Figure \ref{fig:aveSpec100} shows the Sharpe ratio (SR) versus N number of weeks used to construct the moving  average (Eq. (\ref{eq:strategy})) for large range of $N$ . Since we have  6068 weeks (data points) we  present the SR      for our  moving  average  strategy  that  looks  back  up  to  400  weeks  (almost  eight  years).  We also show the SR curves for different weeks. The largest SR are for trades done on Friday and Thursday followed by Wednesdays, Tuesday and Monday. The smallest SR is reserved for trades realized on Monday.  The difference between the weekdays is mostly due to the SR at small N. For instance,          we remind the reader that the Black Monday (October 19, 1987) was the single largest drop in the US      stock market and therefore will have a large effect on weekly trades done on Monday.

Despite the differences of performance between the different weekdays, all graphs show oscillations of the SR as a function of N. We have highlighted in Figure \ref{fig:aveSpec100} areas that are reported in the literature, in particular the literature for stocks, where most work is concentrated, which identifies    three main regions. The first is described in the stock literature as a short-term reversal (Light blue labeled Short-Term in Figure \ref{fig:aveSpec100}) for $N<5$ weeks. The difference with our work is that we do not find a reversal but a continuation (”momentum”) if we look at all DJIA data. However, we know that the problem here is the extension of data used. We already reported that the   autocorrelation of the returns changes from positive to negative and that the positive autocorrelation dominates if we look at all DJIA. Therefore, it is a short-term reversal for data after $1975$, in agreement with the stock literature and a continuation (“momentum”) for data before $1975$. Independent of labelling it reversal or not, we can still label this area as “short-term” and clearly link it to the autocorrelation of the returns.

The second region is normally known as “momentum” or “trend-following” (gray labeled trend- following  in  Figure  \ref{fig:aveSpec100}).  This  is  the  area  where  most  of  the  literature  is  concentrated  and  it  goes from a $N$  of several months to about one year. It also shows the best SR. Based on our studies so            far we expect the performance to be due to a mixture of  autocorrelation  and  drift.  The relative importance between drift and autocorrelation is difficult to  measure  for  non-stationary  data.  For stationary data we found that if we have a well-defined  drift,  the  contribution  of  drift  can  be  as important  as  the  autocorrelation.  Prior studies  look  at  the  average  performance  (not at the  SR  as  we  do) and conclude that the autocorrelation is more important because the drift (which comes in as    square  in  Eq.  \ref{eq:theAve})  must  be much  larger to account  for the  final average performance  of  the  strategy. We also replicate these results. However, since  the  data  is  clearly  not  stationary, we  believe that  using  a formula  derived  under  stationary  assumption  is  at  best  an  approximation.  Moreover,  if  we  take Eq. \ref{eq:theAve} (or even the SR) to apply in this section (all DJIA data without regimes), we  are implicitly    saying  that  any  effect  has  to  be due  to  the  autocorrelation  $\rho$.  In principle  the  autocorrelation  can be a very complex function of N and therefore could  explain  the  SR.  This is  the  only  possible alternative that validates the use of our equation in this section. In other words, our SR equation requires that there is a constant drift for all the $\approx 120$ years of the DJIA and furthermore that         all features of Figure \ref{fig:aveSpec100} are due to a “complex”/”strange” long range autocorrelation. We believe   that this is highly unlikely, especially since Figure  \ref{fig:SRcomp} shows large differences between the stationary analysis performed in the previous section and the one performed in this section.

The third region is known in the literature \cite{Bondt1989} as the long-term reversal (salmon region labeled long-term in Figure \ref{fig:aveSpec100}). Our data shows some reversals since the performance degrade progressively, reaching a minimum SR at around $N \approx 4$ years. To be precise we do not have a full reversal because the SR does not go negative, which would imply that we can “reverse” the sign of the strategy to make money. Therefore, the reversal we report here does not exactly match with the typical effect reported in the stock literature, but we still think that the label is   appropriate. 

\begin{figure}[htbp] \centering
\includegraphics[width=\linewidth]{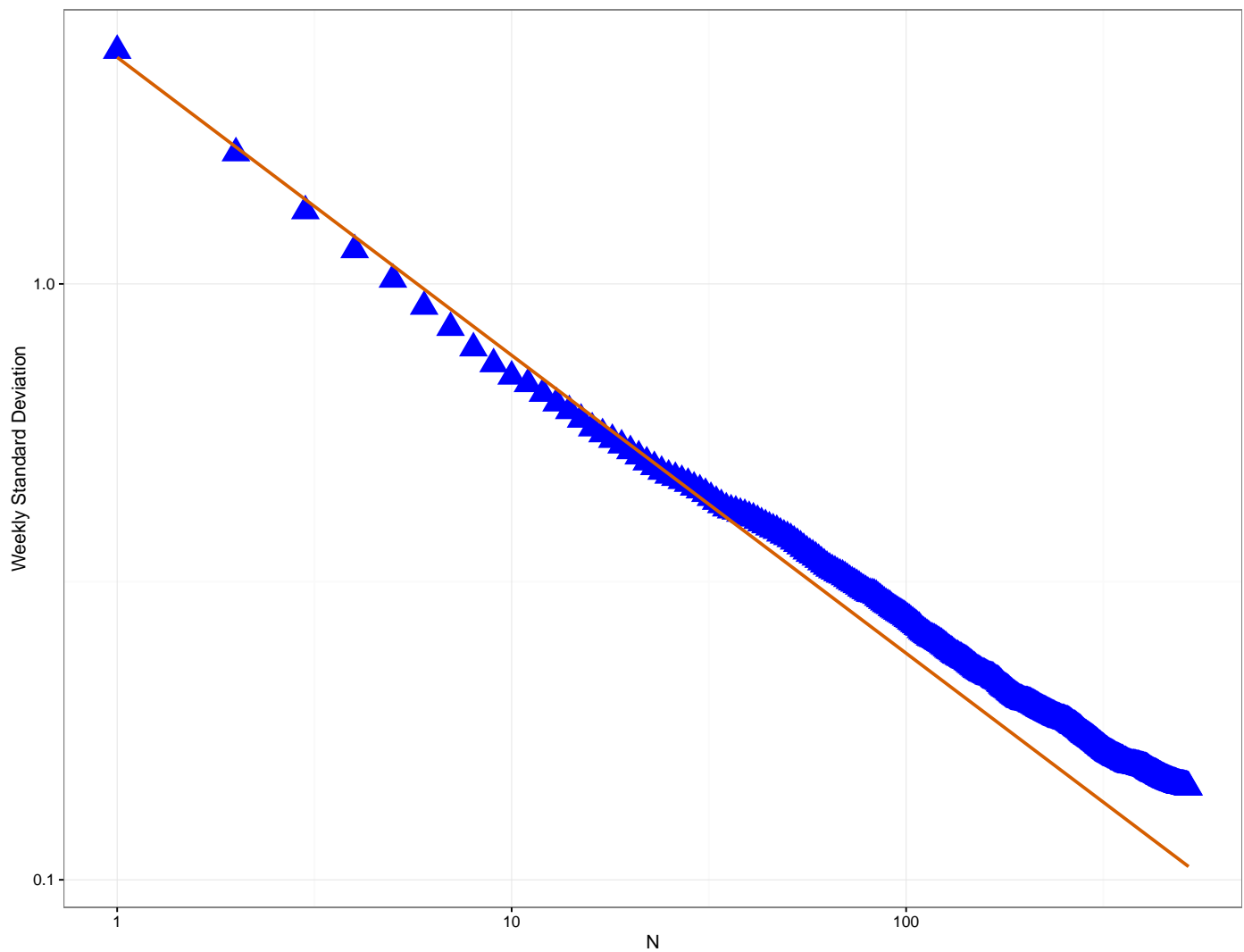} \caption{The
	standard deviation of the strategy applied to all
	DJIA data as a function to the portfolio formation
	period $N$. Data (blue triangles) are shown in a log-log scale to
	illustrate that the standard deviation can be well
	described by a constant times $1/\sqrt{N}$ (solid
	line).
\label{fig:stdFig}} 
\end{figure}

After the third region we find that performance picks up again. In fact, the SR oscillates and,  as       far as we know, this empirical observation has not yet been reported in the literature. Furthermore, the oscillation is due to the oscillation of the average of the strategy (numerator of SR: Eq. \ref{eq:theAve}). The standard deviation of the strategy (denominator of SR: the square root of Eq. \ref{eq:theVar}) is presented in Figure \ref{fig:stdFig}. Notice that despite the complex mathematical formula, the empirical standard deviation can be well approximated by a constant times $1/\sqrt{N}$ which is intuitively expected since the error bar on the mean of independent random numbers is proportional to $1/\sqrt{N}$ .

We  believe that the oscillations for large N  showed in Figure \ref{fig:aveSpec100} are difficult to reconcile with      the stationary theoretical formula for the SR introduced in section \ref{sec:t_model}. It is very difficult to believe (although possible) that the autocorrelation of five years ago will influence the performance  in the current week.
 
 Next we suggest a simple process for the underlying index with oscillating drift that serves as an alternative model and is able to account for the oscillations reported here.
 
 In agreement with our expectations, the curve of the SR dependence on the look-back period is   not qualitatively similar to case I or case II, however we can still use results of the previous section as guide to develop a model.

We learned that both autocorrelation and drift are important to explain the SR of our moving average strategy (\ref{fig:aveSpec}). We  know that case I is more important than case II for large time      lags $N$ (portfolio formation look-back lags) since the SR in case I increases with lag and in case      II it decreases (\ref{fig2}). Considering this, we postulate that the oscillations are mostly due to changes in the drift of case I. That leads us to assume that the theoretical model for the normalized log-returns is given by:

\begin{equation}
\label{eq:nonSmodel}
 r_m(t) = \mu + A\:{\rm sgn}( \sin(2 \pi t/T))+\epsilon,
\end{equation}

\noindent  where $\mu$ stands for the  average  growth  rate,  $A$  for  the  amplitude  of  our  square  wave,  $T$  for  the period and $\epsilon$ for a random variable which we define in the next paragraph.

We take the parameter $\mu$ to be equal to the empirical growth rate of the weekly re-normalize DJIA ($\mu=0.075$ per week). The parameter  $A$ and $T$ are selected to fit the empirical SR oscillation  in Figure 9. We find that T is approximately equal to 3.5 years (180 weeks) and that $A\approx 2\times\mu$. The root-mean-square σ of the random variable $\epsilon$ is equal to the empirical root-mean-square of the re-normalized DJIA weekly returns over the entire period ($\sigma \approx 2\times\mu /week$).

The model in Eq.  (\ref{eq:nonSmodel})  shows  the  long  term  reversals  reported  by  \cite{Bondt1985,Bondt1987,Bondt1989,Fama} for stocks and  indices.  The  period  of  3.5  years  implies  a  mean reversion of about 1.75 years which is within the prior reported value of 1.5 to 5 years \cite{Bondt1989}.  In  terms  of  correlations,  our  model  has  a  small  positive  autocorrelation  that  progressively goes negative and back up again: oscillating with the period $T$ even when $\epsilon$ is independent and identically distributed  random  variable.  However, since  the  empirical  random  variable level is between 7 to 20 times larger than $\mu \pm A$, such correlations are not easy to detect by calculating the autocorrelation function. 

In order to compute the SR versus $N$ for our theoretical model (Eq. (\ref{eq:nonSmodel})), we perform Monte Carlo (MC) simulations (500 of them). We consider two types of $\epsilon$. The first one is an IID Gaussian random variable with zero drift and variance equal to the empirical variance of the re-normalized data. This is the simplest case and assumes that all the performance for a momentum strategy is only due to the drift ($\mu$). The second one is a MA process with zero mean and variance    as the Gaussian IID random variable, but with best fit positive autocorrelation for lag 1 (0.05) and   lag 20 (0.08). The second model takes in consideration that short lags might be strongly influenced    by autocorrelation (case II situation) and therefore both case I and case II co-exist as in Figure \ref{fig:aveSpec100}. The MA process is selected by first choosing the non-zero autocorrelation lags and then by fitting the model to the empirical data ($SR$ vs $N$ Friday to Friday) to find the numerical coefficients, which imply a correlation of 0.05 and 0.08 for lag 1 and lag 20. We select the non-zero lags by using the intuition developed for the stationary model and by keeping the number of non-zero lags minimal. From Eq. (\ref{eq:case2}) and Figure \ref{fig2}, we know that a MA model generates a localized hump in the graph  $SR$ versus $N$ in the case of stationary data. The location of the hump is centered at the location of the autocorrelation. Assuming that this is approximately valid here, the SR vs N for uncorrelated $\epsilon$ can be shifted up in the most relevant places to get a better fit. Therefore, by looking at the empirical SR vs N of Figure \ref{fig:aveSpec100}, we postulate that the most relevant lags should be around lag 1 and lag 20, since we need to shift the theoretical $SR$ vs $N$ generated with $\epsilon$ being IID Gaussian (thick dashed red) at least at lag 1 and lag 20, to produce the $SR$ vs $N$ from the MA generated $\epsilon$  (thick dashed black).

Figure \ref{fig:aveSpec100} shows the empirical $SR$ versus
$N$  (thin solid lines) and 2 theoretical lines (thick   dashed red  and  black).  The  best  fit  is  achieved  with  s  drawn  form  a  MA process  (dashed  black  line). The  worst  fit  is  with  $\epsilon$ drawn  from  an  IID Gaussian  distribution (red  dashed  line).  The oscillations  are well captured by the uncorrelated $\epsilon$, especially for very large $N$ , however it underestimates the SR ratio for small $N$. We have shown in Figure \ref{fig6} that before 1975 the autocorrelation is significantly positive and, since we are in effect averaging the data  by  treating  the  100  years  of  the  DJIA  as stationary, we need to include large positive short range autocorrelation to account for data before 1975. The result is the much better fit of the dashed black line in Figure \ref{fig:aveSpec100}.

Based on Figure \ref{fig:aveSpec} one could expect to find a small autocorrelation effect on the SR for  $N \approx 20$, which is not what we find by fitting the SR here. However, this is misleading, since we would be comparing approximately stationary data to non-stationary data. Furthermore, even if some regimes have “large” lag 20 autocorrelation, this autocorrelation would have been averaged out.   That is, the difference between Figure \ref{fig:aveSpec} and Figure \ref{fig:aveSpec100} is that Figure \ref{fig:aveSpec} is constructed by taking an average over all the SR versus $N$ curves created within our stationary patches. In other words, all SR for $N  \approx 20$ that have a hump are averaged with all SR which do not show a hump. Since at $N \approx 20$ the contribution to SR of the autocorrelation hump is of order $1/\sqrt{20}$ and the contribution of drift is of order 1 (Eq. (\ref{eq:theVar})), the average is dominated by the drift. The result is that the effect of the autocorrelation for $N$ larger than 15 in Figure \ref{fig:aveSpec} should not be evident.

\begin{figure}[htbp] \centering
\includegraphics[width=\linewidth]{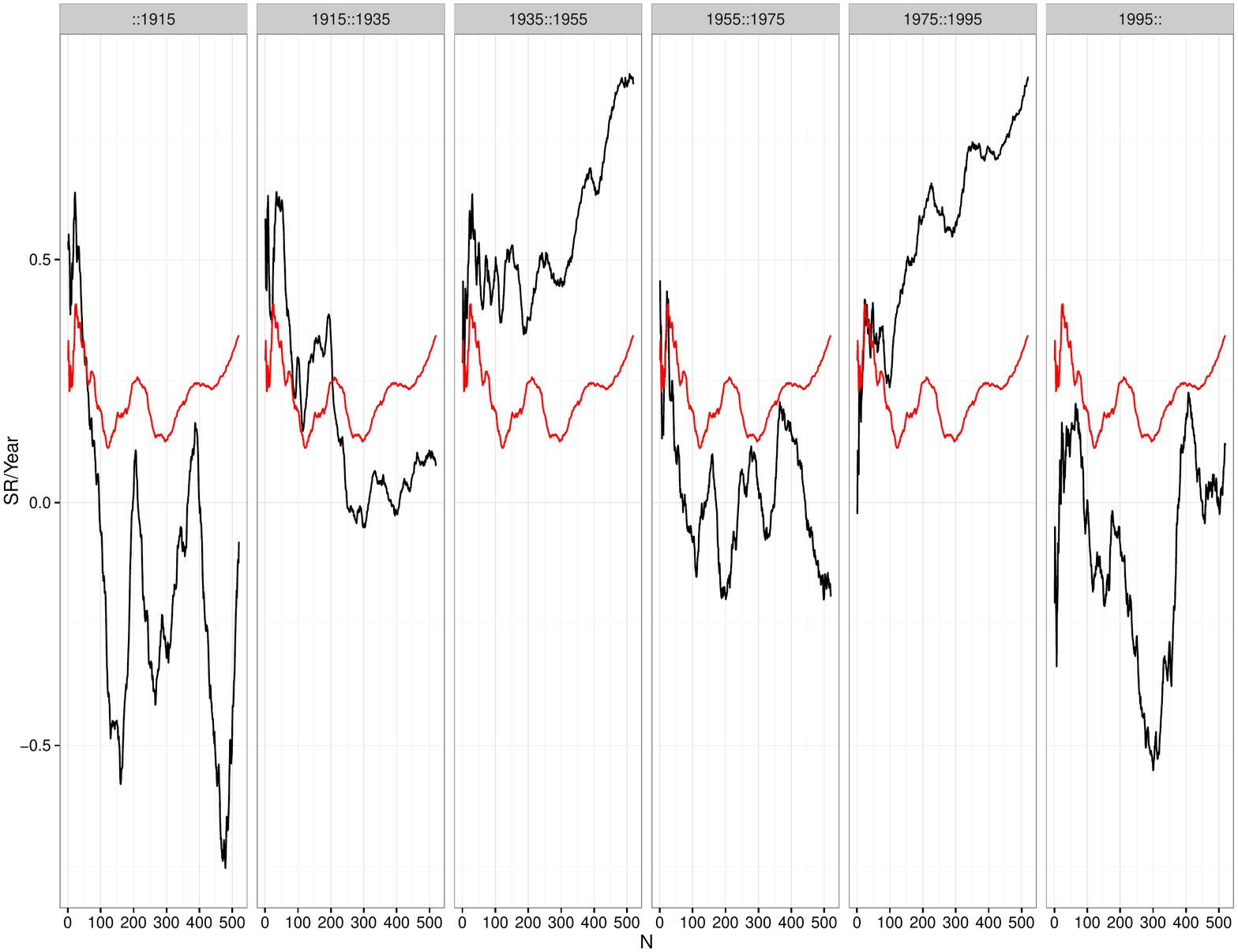}
\caption{SR vs. portfolio formation period $N$ (weeks) for the strategy (Eq. \ref{eq:strategy}) per 20 year sub-periods. The solid red   line refers to the strategy applied to 119 years of the DJIA while the solid black line refers to the strategy applied within a 20 year  period.
\label{fig:subPer}} 
\end{figure}

Figure \ref{fig:subPer} shows the SR vs $N$  for 20 years sub-periods of the nearly 120 years of the DJIA. Each panel show the average SR versus N of portfolio formation periods over five weeks (Monday -  Monday to Friday - Friday). The red line results by applying our strategy over the 119 years of data and the black curve is the result of applying the strategy within each sub-period. The graph illustrates the stability of the oscillations. We see that all 20 year sub-periods show oscillations. However, the period of such oscillations can change (1955 to 1975) and/or there might be a strong trend as a function of N (1915 to 1935 and in particular 1975 to 1995). Therefore, the graph (Figure \ref{fig:subPer}) illustrates that the oscillations which alternate reversions with trend-following for large N (N $> 4$ years) is present for all sub-periods even though quantitatively different parameters might be needed to describe each sub-period.

Finally, we would like to close with a brief statement about the performance of our trading rule from a practical point of view. We emphasize again that we are not proposing a new trading rule and that our goal is to understand the performance of a simple and representative rule which has been evoked many times before. Nevertheless, for the sake of clarity and completeness, we compare our trading rule with the buy-and-hold portfolio. If the investor had bought the DJIA index   and
done nothing for 120 years, that investor would have had a Sharpe ratio of $\approx 0.26$ (average return
of about 4\% and a volatility of 19\% per year). Figure 9 shows that our strategy would have
resulted in a Sharpe ratio between $0.35$ to $0.45$ if the same investor would have selected N  within      the trend-following range (gray region in the Figure \ref{fig:aveSpec100}). For example, if the investor selects $N=25$    weeks, the return of our simple rule is approximately 9\% for the same average volatility of 19\% per year. Therefore, this trading rule could show potential from a practical  point  of  view,  if transaction costs are ignored\footnote{Transactions costs are a very important issue that has to be addressed whenever a high turnover strategy is advocated as superior. Therefore, our comparison here is merely hypothetical.}. We are sure that given the simplicity of our strategy, the investor should do much better. See, for instance, the work of \cite{Papailias2015} for examples of other rules. What  remains  an  open  question  is  whether  the  same  underlying  mechanism  presented  in  Equation \ref{eq:theAve}, Equation \ref{eq:theVar} and model  \ref{eq:nonSmodel} is relevant for the many different rules and strategies proposed in the literature. That is, whether our simple rule is a good first order representation of most price based trading strategies that use moving averages.

\section{Conclusions}

In the present work we perform a statistical analysis on a moving average strategy and find a closed      form formula for the expected value and the variance. We are therefore able to present an analytic expression for the Sharpe ratio. The  innovation  here  allows  one  to  compute  and  discuss  the  risk adjusted  performance  of  strategies  that  use  past  performance,  in  particular  past  returns  to  decide how  to trade.

The theory developed here assumes that the time series have stationary patches. One key point        is to find the proper stationary patch. Once this condition is satisfied, we get good agreement between theory and data as discussed in section \ref{sec:E_stationary}. Nevertheless, when a longer time series is considered, the stationary assumption may no longer be valid. In this scenario, we proposed a stochastic model in section \ref{sec:NE_stationary}. This model agrees well with the data by introducing a periodic change in the average growth rate of the DJIA

In section \ref{sec:E_stationary} we  find that both autocorrelation and drift can be important in describing the       risk adjusted performance of a time series momentum strategy. More originally, we show that the Sharpe ratio appears to present two phases. The first phase is for short look-back periods and the second one is for long look-back periods. The first phase is mainly driven by autocorrelation while the second phase is driven by the drift (Figure \ref{fig:aveSpec}). Furthermore, there is a behavioral change in 1975, where the first lag autocorrelation of weekly returns go from mainly positive before 1975 to mainly negative after 1975

In section \ref{sec:NE_stationary}, we find an oscillatory SR for long portfolio formation periods which we associate     with periodic cycles in the market (Figure \ref{fig:aveSpec100}). We emphasize  that  the  results  in  section  \ref{sec:NE_stationary}  are inherently unstable  since  we  use  119  years  of  DJIA  data.  It  means  that  if  we  repeat  the  same study for different sub-samples we will find different quantitative and sometimes qualitative results (Figure \ref{fig:subPer}). Nevertheless the oscillations seem to indicate  that  economic  cycles  are  substantially important to the performance of our trading rule.

\newpage

\appendix
\section{Mathematical derivations}
\label{sec:app1}

The characteristic function for the multivariate Gaussian random variable is

\begin{equation}
\phi(s) =  \exp\Big( i\sum_{i} s_{i} \mu_{i} -\frac{1}{2} \sum_{i} V_{ii} s_{i}^{2} + \frac{1}{2} \sum_{i \neq j} s_{i} s_{j} V_{ij}\Big ).
\end{equation}
We will calculate each term of Eq. (\ref{eq:risk2}) separately. The summand of the first term is
\begin{equation}
\label{eq:riskA1}
\left \langle X_t^{2} X_{t-i}^{2}\right \rangle =\left \langle X_{k}^{2} X_{i}^{2}\right \rangle = \frac{\partial^{2}}{\partial s_{k}^{2}} \frac{\partial^{2}}{\partial s_{i}^{2}} \phi(s) |_{s=0}
\end{equation}
where we have simplified our notations by taking current time $t$ as $k$; and
past times being labeled as $i$, $j$ or $q$. The double partial derivatives in (\ref{eq:riskA1}) result into:

\begin{eqnarray}
\label{eq:cal1}
V_{ii} V_{kk} \phi(s) \\ \nonumber
 -V_{ii} [i \mu_{k}-V_{kk} s_{k}-\sum_{j} s_{j}V_{ij}]^{2} \phi(s) \\ \nonumber
+V_{ki}^{2} \phi(s) \\ \nonumber
-V_{ki}
 [i\mu_{k}-V_{kk}s_{k}-\sum_{j}s_{j}V_{kj}][i\mu_{i}-V_{ii}s_{i}-\sum_{j} s_{j}V_{ij}]\phi(s) \\ \nonumber
+V_{ki}^{2} \phi(s) \\ \nonumber
-V_{ki}
 [i\mu_{k}-V_{kk}s_{k}-\sum_{j} s_{j} V_{kj}][i\mu_{i}-V_{ii}s_{i}-\sum_{j} s_{j}V_{ij}]\phi(s) \\ \nonumber
-V_{ki}
 [i\mu_{k}-V_{kk}s_{k}-\sum_{j} s_{j} V_{kj}][i\mu_{i}-V_{ii}s_{i}-\sum_{j} s_{j}V_{ij}]\phi(s) \\ \nonumber
-V_{ki}
 [i\mu_{k}-V_{kk}s_{k}-\sum_{j} s_{j} V_{kj}][i\mu_{i}-V_{ii}s_{i}-\sum_{j} s_{j}V_{ij}]\phi(s) \\ \nonumber
-V_{kk}
[i\mu_{i}-V_{ii} s_{i}-\sum_{j} s_{j} V_{ij}]^{2} \phi(s) \\ \nonumber
+[i \mu_{k} - V_{kk} s_{k} - \sum_{j} s_{j} V_{kj}]^{2} [i \mu_{i} - V_{ii} s_{i} - \sum_{j} s_{j} V_{ij}]^{2} \phi(s) \nonumber
\end{eqnarray}

\noindent where $j$ is the index of  the summation. Let $s=0$, thus
Eq. (\ref{eq:cal1}) is reduced to:

\begin{equation}
\left \langle X_{k}^{2} X_{i}^{2}\right \rangle=V_{ii}V_{kk}+2V_{ki}^{2}+V_{ii}\mu_{k}^{2}+4V_{ki}\mu_{i}\mu_{k}
+\mu_{i}^{2}V_{kk}+\mu_{i}^{2}\mu_k^{2}
\end{equation}

\noindent Now, the summand of the second term of Eq. (\ref{eq:risk2}) is given by

\begin{equation}
\label{eq:riskA2}
\left \langle X_t^2X_{t-i}X_{t-j}\right \rangle = \left \langle X_{k}^{2} X_{i}X_{j}\right \rangle = \\
\frac{\partial^{2}}{\partial s_{k}^{2}} \frac{\partial}{\partial s_{i}} \frac{\partial}{\partial s_{j}} \phi(s) |_{s=0}
\end{equation}
Notice that we use $j$ to indicate a different time lag in Eq. (\ref{eq:riskA2}). The partial derivatives in Eq. (\ref{eq:riskA2}) result into:

\begin{eqnarray}
\label{eq:cal2}
+V_{ij}V_{kk}\phi(s)\\ \nonumber
-V_{ij} [i \mu_{k} - V_{kk} s_{k} - \sum_{q} s_{q} V_{kq}]^{2} \phi(s) \\ \nonumber
+ V_{ki}V_{kj} \phi(s) \\ \nonumber
-V_{ki} [i \mu_{k} - V_{kk} s_{k} - \sum_{q} s_{q} V_{kq}] [i \mu_{j} - V_{jj} s_{j} - \sum_{q} s_{q} V_{jq}] \phi(s) \\ \nonumber
+V_{ki}V_{kj} \phi(s)\\ \nonumber
-V_{kj} [i \mu_{k} - V_{kk} s_{k} - \sum_{q} s_{q} V_{kq}] [i \mu_{i} - V_{ii} s_{i} - \sum_{q} s_{q}V_{iq}] \phi(s) \\ \nonumber
-V_{kj} [i \mu_{k} - V_{kk} s_{k} - \sum_{q} s_{q} V_{kq}] [i \mu_{i} - V_{ii} s_{i} - \sum_{q} s_{q}V_{iq}] \phi(s) \\ \nonumber
-V_{ki} [i \mu_{k} - V_{kk} s_{k} - \sum_{q} s_{q} V_{kq}] [i \mu_{j} - V_{jj} s_{j} - \sum_{q} s_{q} V_{jq}] \phi(s) \\ \nonumber
-V_{kk} [i \mu_{i} - V_{ii} s_{i} - \sum_{q} s_{q} V_{iq}] [i \mu_{j} - V_{jj} s_{j} - \sum_{q} s_{q} V_{jq}] \phi(s) \\ \nonumber
+ [i \mu_{k} - V_{kk} s_{k} - \sum_{q} s_{q} V_{kq}]^{2} [i \mu_{j} - V_{jj} s_{j} - \sum_{q} s_{q} V_{jq}] \\ \nonumber
\times [i \mu_{i} - V_{ii} s_{i} - \sum_{q} s_{q} V_{iq}]  \phi(s) \\ \nonumber
\end{eqnarray}

\noindent Let $s=0$, then (\ref{eq:cal2}) becomes

\begin{equation}
V_{ij}V_{kk}+2V_{ki}V_{kj}+2V_{kj}\mu_{i}\mu_{k}
+V_{ij}\mu_{k}^{2}+2V_{ki}\mu_{j}\mu_{k}
+V_{kk}\mu_{j}\mu_{i}+\mu_{k}^{2} \mu_{i} \mu_{j}
\end{equation}

\noindent Hence, the first term of (\ref{eq:risk1}), i.e. the square of strategy return ($R$), is given by

\begin{eqnarray}
\label{eq:cal3}
\left \langle R^{2} \right \rangle &=& \frac{1}{N^2}\Big(
\sum_{i} \big(V_{ii}V_{kk}+2V_{ki}^{2}+V_{ii}\mu_{k}^{2}+4V_{ki}\mu_{i}\mu_{k}
+\mu_{i}^{2}V_{kk}+\mu_{i}^{2}\mu_k^{2}\big)\nonumber \\ 
&+&\sum_{i,j, i \neq j} \big(V_{ij}V_{kk}+2V_{ki}V_{kj}+2V_{kj}\mu_{i}\mu_{k}
+V_{ij}\mu_{k}^{2}+2V_{ki}\mu_{j}\mu_{k}+  
V_{kk}\mu_{j}\mu_{i}+\mu_{k}^{2} \mu_{i} \mu_{j}\big)\Big). 
\end{eqnarray}

\noindent Now,  we compute the second term of Eq. (\ref{eq:risk1}), i.e. the square of the average, and we get

\begin{eqnarray}
\label{eq:cal4}
\left \langle R \right \rangle^{2} &=& \frac{1}{N^2}\Big(\sum_{i} \big(V_{ij} +\mu_{i}\mu_{j}\big)\Big)^{2} \nonumber \\&=&
\frac{1}{N^2}\Big(\sum_{i} \big(V_{ki}^{2} + 2V_{ki}\mu_{k}\mu_{i}+\mu_{k}^{2}\mu_{i}^{2}\big)  \nonumber \\
&+&\sum_{i,q,i\neq q} \big(V_{ki}V_{kq}+V_{ki}\mu_{k}\mu_{q}+V_{kq}\mu_{k}\mu_{i}
+\mu_{k}^{2}\mu_{i}\mu_{q}\big)\Big)
\end{eqnarray}
\noindent where we use the same simplifying notation ($i$,$j$,$k$) to
represent the time lags.

Finally, from  Equations (\ref{eq:cal3}) and (\ref{eq:cal4}), we obtain Eq. (\ref{eq:risk1})as the following:

\begin{eqnarray}
\label{eq:finalGeneralVar}
Var(R) &=& \frac{1}{N^2}\Big(\sum_{i} \big(V_{ii}V_{kk}+V_{ki}^{2}+V_{ii}\mu_{k}^{2}+2V_{ki}\mu_{i}\mu_{k}+V_{kk}\mu_{i}^{2}\big) \\ \nonumber
&+&\sum_{i,j,i\neq j} \big(V_{ij}V_{kk}+V_{ki}V_{kj}+V_{kj}\mu_{i}\mu_{k}+V_{ij}\mu_{k}^{2}+V_{kk}\mu_{i}\mu_{j} +V_{ki}\mu_j\mu_k\big)\Big). \nonumber
\end{eqnarray}
Now, assuming  that the variance $V$ and the drift $\mu$ are constants and that the autocorrelation depends only on time lag, (\ref{eq:finalGeneralVar}) can be converted to Eq. (\ref{eq:theVar}), i.e.

\begin{eqnarray}
\label{eq:theVarApp} Var(R) &=& \frac{1}{N^{2}}\bigg[ N V^{2} + N \mu^{2} V
+ N^{2} V \mu^{2} \\ \nonumber &+&V^{2} \big(\sum_{i=1}^{N} \rho(t,t-i)\big)^{2} +V^{2}\sum_{i,j=1,i\neq
j}^{N} \rho(t-i,t-j)\\
\nonumber &+&\mu^{2} V \Big(2\sum_{i=1}^{N} \rho(t,t-i) +\sum_{i,j=1,i\neq
j}^{N}\big(\rho(t,t-j)+\rho(t-i,t-j)+\rho(t,t-i)\big)\Big) \bigg]  
\end{eqnarray}

\noindent where we convert back to the time lag notation used in the main text by reverting the $i,j,k$ notation. For example, we take $V_{ki} = V \rho(t,t-i)$,$V_{ij} = V \rho(t-i,t-j)$ and $V_{kk}=V$.

\newpage

\section{SR and return normalization}
\label{sec:app2}

\begin{figure}[htbp] \centering
\includegraphics[width=\linewidth]{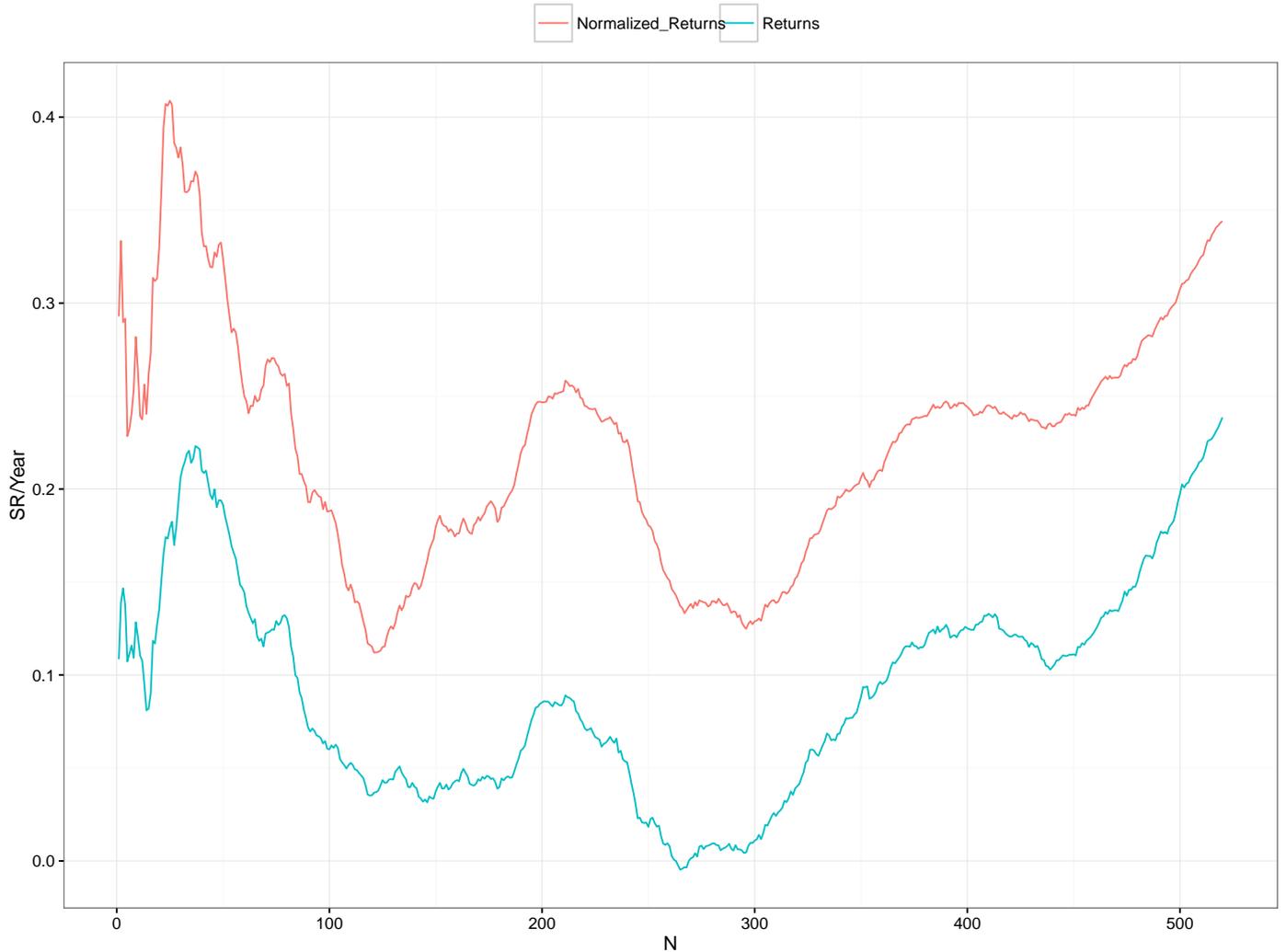} \caption{SR for strategy (Eq. \ref{eq:strategy} as a function of $N$ when applied to log-returns that are scaled/normalized by a measure of the local standard deviation (Eq. \ref{eq:normalizeData}) [top red line] compared to the same strategy applied to log-returns without normalization (bottom blue line).\label{fig:app2}}
\end{figure}

How  important  is  normalization for Eq.  \ref{eq:normalizeData}?  Figure  B1  shows  the  SR  versus  portfolio  formation  \ref{fig:app2} for weekly  DJIA log-returns. The  upper  red curve shows  the SR for  normalized log-returns using    Eq.  \ref{eq:normalizeData}  as  described  in  the  empirical  section  of  the  article.  The  lower  blue  curve  shows  the  SR for  not-normalized  data.  The  most  relevant  features  which  are  the  SR  oscillations  are  still  present and in sync for the normalized and not-normalized data. The figure shows that the normalization generates a better SR for each $N$  but it does not substantially change how  SR changes as a function      of $N$  which is the main topic of our   study.

\newpage

\section{World indices}
\label{sec:app3}

This section shows that the results (primarily oscillations of SR versus N) discussed in the empirical section of this article are still true for most developed markets. We  have  only limited data for non-      US markets, therefore we show results for different sub-periods depending on the availability of the   data.

\begin{figure}[htbp] \centering
\includegraphics[width=\linewidth]{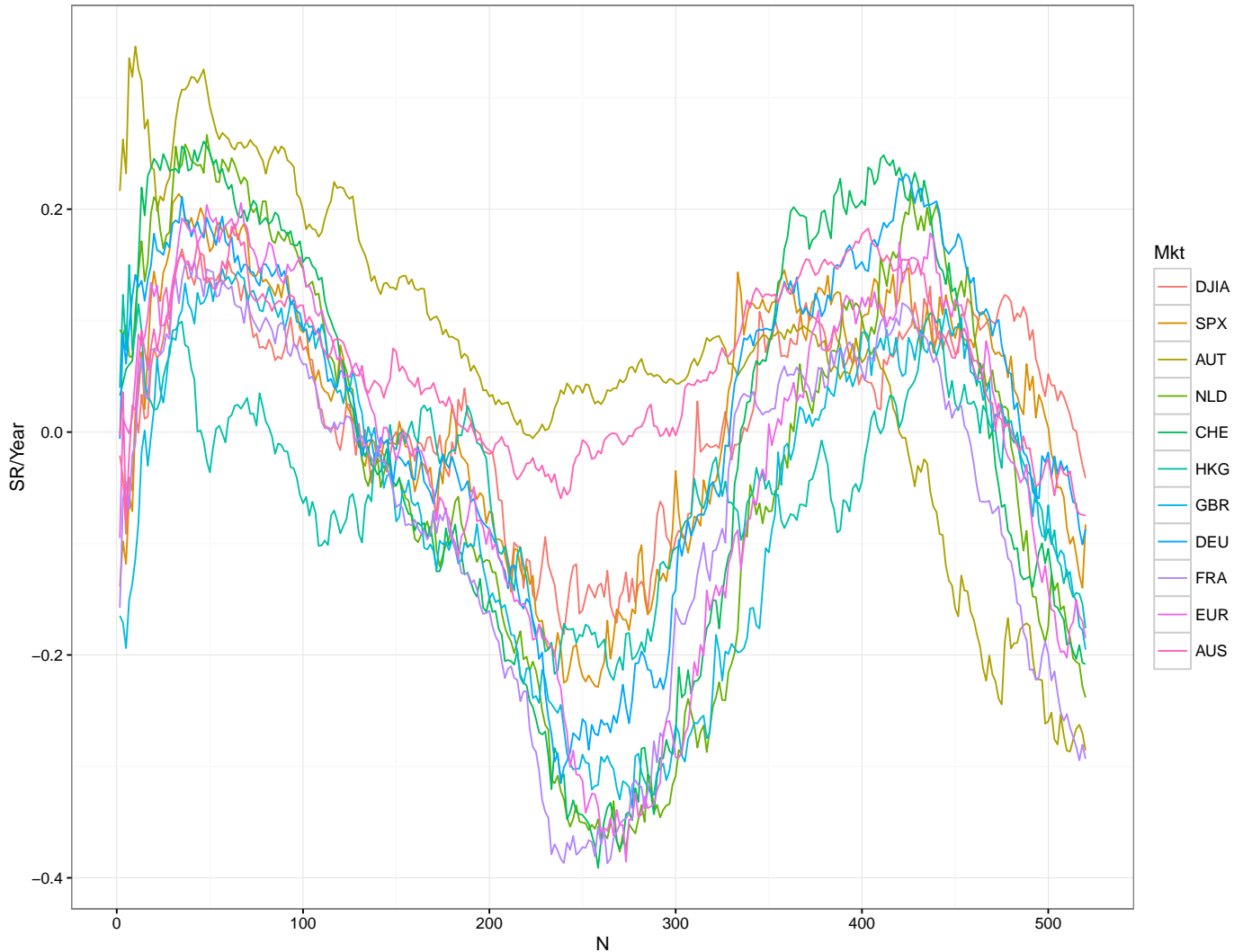}
\caption{SR for strategy (Eq. \ref{eq:strategy}) applied to 11 developed world markets from 1994 to 2016 as a function of portfolio creation look-back N weeks. Indices are labelled by the abbreviations and refer to the largest indices for each country. For instance, Germany (DEU) we show the SR of our strategy applied to the DAX. The SR for most markets are very close, the exception is Japan (JPN) and Australia (AUS).\label{fig:app3}}
\end{figure}

In particular the sub-period after 1994 shows  a  remarkable  agreement  between  the  SR  for  11 different  developed  equity  indices  (Figure  \ref{fig:app3}).  Figure  \ref{fig:app3} illustrates  that  the  oscillation  is  present in all markets despite few  differences

\begin{figure}[htbp] \centering
\includegraphics[width=\linewidth]{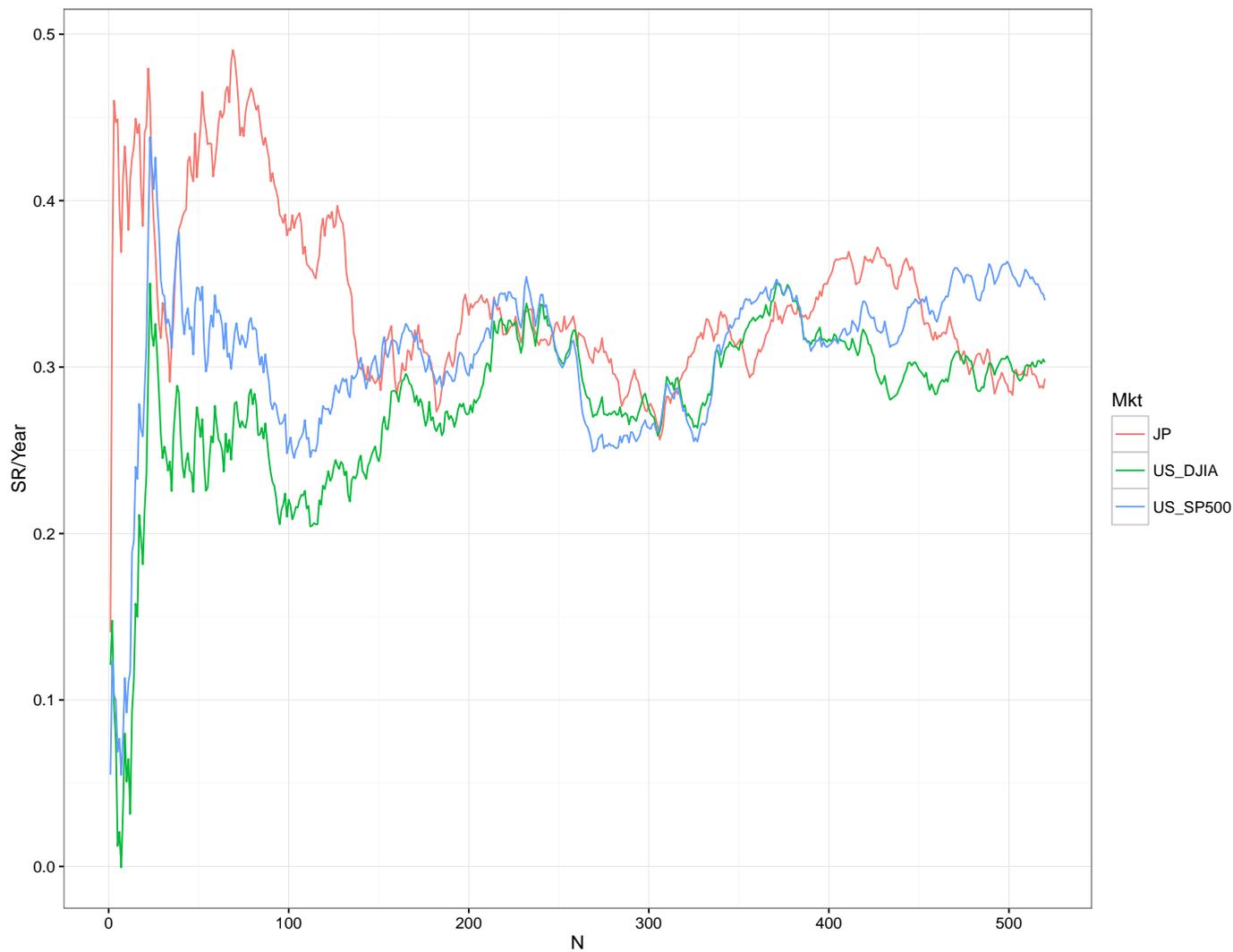} \caption{Figure C2. SR for strategy (Eq.\ref{eq:strategy}) applied to three indices from 1950 to 2016 as a function of N weeks. For JP we use the Nikkei 225.\label{fig:app31}}
\end{figure}

The differences in the SR as a function of N is more pronounced for returns that start in 1950.   Due to data availability we show the SR for only 3 indices: Nikkei 225 (JPN), the DJIA and   the S\&P 500. Figure \ref{fig:app31} shows large differences between JPN and the USA up to $N \approx 2$ years. After that the indices oscillate together. However, the main features (i.e. oscillations) are  qualitatively similar to for all three indices.

\newpage

\section{SR and trading frequency}
\label{sec:app4}

\begin{figure}[htbp] \centering
\includegraphics[width=\linewidth]{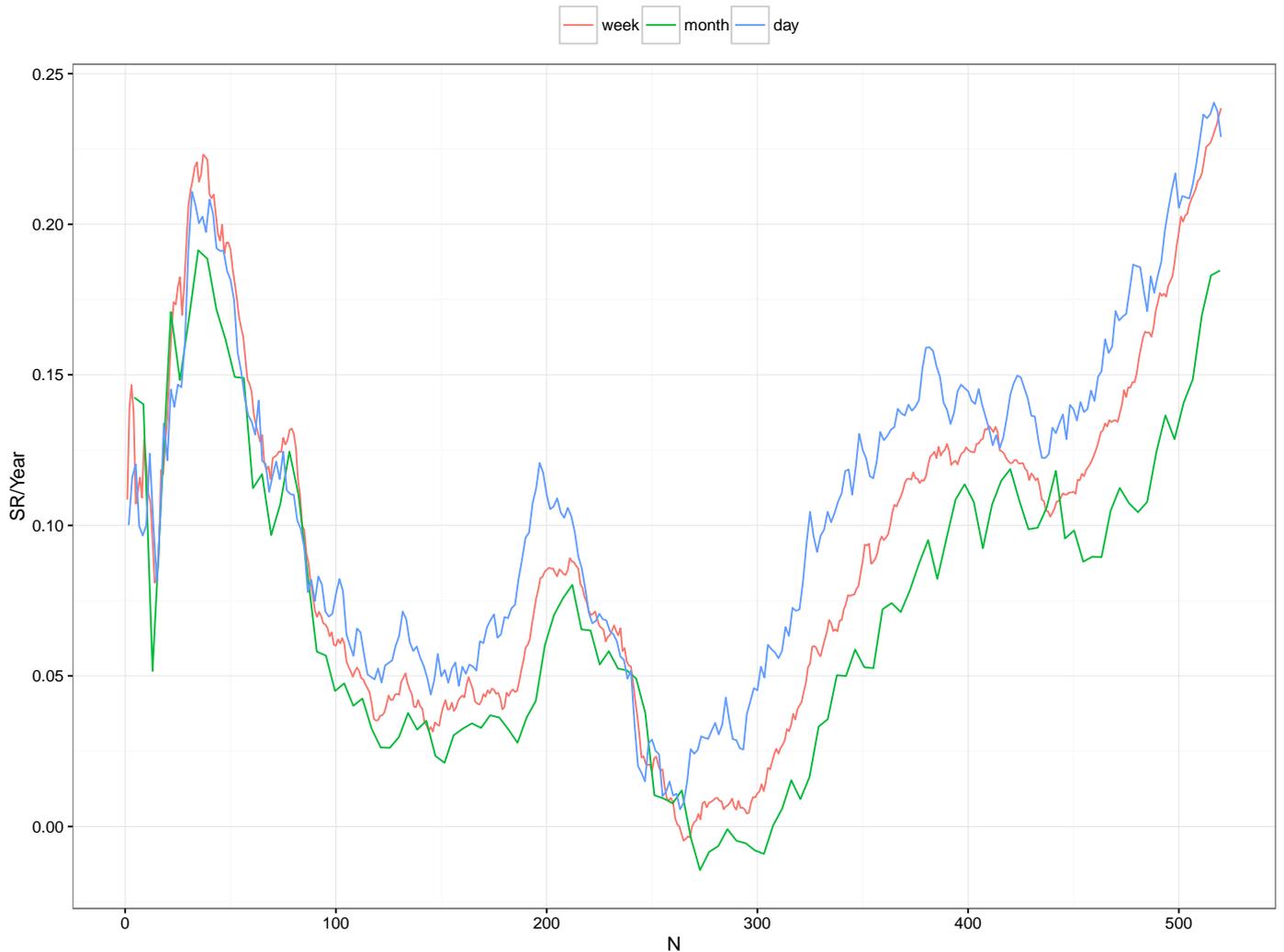} \caption{SR vs $N$ (weeks) trading at different frequency. In order to convert daily $N$ to weekly $N$ and to have both of them plotted   on the same graph, we assume that one week has 6 days. In order to convert monthly $N$ into weekly $N$ we assume that one month has 4.3  weeks.
\label{fig:app4}} 
\end{figure}

How important is to trade weekly? Are the oscillations or any other feature mostly due to weekly    trade? Figure \ref{fig:app4} shows the Sharpe ratio  (SR)  as  a  function  of  the  portfolio  formation  period applying  our  strategy  (Eq.  \ref{eq:strategy})  daily,  weekly  or  monthly.  An  important  distinction  is  that  we  apply the strategy (Eq. \ref{eq:strategy}) using non-normalized log-returns. That makes the comparison simple since the normalization using daily, weekly or monthly returns is difficult to define in order for the different  trading  frequencies  to  be  compatible.  Nevertheless,  the  effect  of  normalization  is  negligible  here (see section \ref{sec:app2} for the effect of   normalization).

Figure \ref{fig:app4} shows that the SR for trading daily, weekly or monthly overlaps. That is, the results     of this study do not depend on the trading frequency.

From a practical point of view, trading daily is much more complex and costly. However, trading cost effects are ignored at any point in this study, but clearly should be accounted in any real life implementation.

\newpage

\section*{Acknowledgments}

ACS would like to thank the participants of R/Finance 2013 at UIC and QCMC 2015 at the         Max Planck Institute for comments and the organizers for financial support to present preliminary versions of this paper. FFF acknowledges financial support from Fundac¸a˜o Amparo \`a Pesquisa do Estado de S\~ao Paulo (FAPESP) and Funda\c{c}\~ao Instituto de F\'isica Te\'orica (FIFT) for hospitality.   We thank Constantin Unanian for detailed comments. J-YY is grateful to Academia Sinica Institute of Mathematics (Taipei, Taiwan) for their hospitality and support during some extended visit.

\bibliographystyle{plain}
\bibliography{trendF}

\end{document}